\pdftrailerid{}
\documentclass[conference]{IEEEtran}
\IEEEoverridecommandlockouts

\usepackage[T1]{fontenc}
\usepackage[utf8]{inputenc}
\usepackage{amsmath}
\usepackage{amssymb}
\usepackage{array}
\usepackage{booktabs}
\usepackage{cite}
\usepackage{graphicx}
\usepackage{xcolor}
\usepackage[hidelinks]{hyperref}
\usepackage{url}

\graphicspath{{figures/}}
\newcommand{\ActiveTwinPairCount}{450}
\newcommand{\CVFoldCount}{5}
\newcommand{\ConfidenceLevel}{95}
\newcommand{\DirectTwinAUROCMax}{0.995}
\newcommand{\DirectTwinAUROCMin}{0.980}
\newcommand{\DirectTwinBTEMax}{0.114}
\newcommand{\DirectTwinBTEMin}{0.036}
\newcommand{\FixedTwinFBRMax}{0.986}
\newcommand{\FixedTwinFBRMin}{0.686}
\newcommand{\FixedTwinTPRMax}{1.000}
\newcommand{\FixedTwinTPRMin}{0.943}
\newcommand{\LocalCovarianceAUROCMax}{0.964}
\newcommand{\LocalCovarianceAUROCMin}{0.943}
\newcommand{\LocalCovarianceBTEMax}{0.136}
\newcommand{\LocalCovarianceBTEMin}{0.107}
\newcommand{\OracleHeadAUROCMax}{0.984}
\newcommand{\OracleHeadAUROCMin}{0.979}
\newcommand{\PrimaryCVBTEMax}{0.329}
\newcommand{\PrimaryCVBTEMin}{0.229}
\newcommand{\PrimaryDeltaAUROCMax}{0.297}
\newcommand{\PrimaryDeltaAUROCMin}{0.139}
\newcommand{\PrimaryPairRankingMax}{0.929}
\newcommand{\PrimaryPairRankingMin}{0.829}
\newcommand{\PrimaryTwinAUROCMax}{0.860}
\newcommand{\PrimaryTwinAUROCMin}{0.701}
\newcommand{\PrimaryTwinPairCount}{70}
\newcommand{\PrimaryWorstGroupCategoryN}{7}
\newcommand{\ReferenceSecondaryFBRMax}{0.140}
\newcommand{\ReferenceSecondaryFBRMin}{0.050}
\newcommand{\ReferenceSecondaryTPRMax}{0.907}
\newcommand{\ReferenceSecondaryTPRMin}{0.562}
\newcommand{\ReferenceTwinFBRMax}{0.257}
\newcommand{\ReferenceTwinFBRMin}{0.129}
\newcommand{\ReferenceTwinTPRMax}{0.971}
\newcommand{\ReferenceTwinTPRMin}{0.871}
\newcommand{\ResampleDrawCount}{10{,}000}
\newcommand{\SecondJudgeAllFieldSwapConsistency}{0.891}
\newcommand{\SecondJudgeArmCount}{900}
\newcommand{\SecondJudgeBinaryKappa}{0.958}
\newcommand{\SecondJudgeBinaryKappaLower}{0.939}
\newcommand{\SecondJudgeBinaryKappaUpper}{0.975}
\newcommand{\SecondJudgeBlockedRate}{0.006}
\newcommand{\SecondJudgeConsensusMaxCVBTEShift}{0.0165}
\newcommand{\SecondJudgeConsensusMaxFBRShift}{0.0143}
\newcommand{\SecondJudgeConsensusMaxPairShift}{0.0199}
\newcommand{\SecondJudgeConsensusMaxTPRShift}{0.0173}
\newcommand{\SecondJudgeConsensusMaxTwinShift}{0.0231}
\newcommand{\SecondJudgeConsensusPairCount}{432}
\newcommand{\SecondJudgeExactlyOneAgreement}{0.967}
\newcommand{\SecondJudgeHarmfulAgreement}{0.973}
\newcommand{\SecondJudgeHarmlessAgreement}{0.911}
\newcommand{\SecondJudgeOrientationAgreement}{0.992}
\newcommand{\SecondJudgeOrientationSwapConsistency}{0.993}
\newcommand{\SecondJudgePairCount}{450}
\newcommand{\SecondJudgePairValidityAgreement}{0.992}
\newcommand{\SecondJudgePresentationCount}{900}
\newcommand{\SecondJudgeRawAgreement}{0.942}
\newcommand{\SecondJudgeRawAgreementLower}{0.927}
\newcommand{\SecondJudgeRawAgreementPercent}{94.2\%}
\newcommand{\SecondJudgeRawAgreementUpper}{0.958}
\newcommand{\SecondJudgeReadoutCount}{7}
\newcommand{\SecondJudgeSameFrameAgreement}{0.721}
\newcommand{\SecondJudgeSameTopicAgreement}{0.929}
\newcommand{\SecondJudgeStrictMaxCVBTEShift}{0.0250}
\newcommand{\SecondJudgeStrictMaxFBRShift}{0.0361}
\newcommand{\SecondJudgeStrictMaxPairShift}{0.0333}
\newcommand{\SecondJudgeStrictMaxTPRShift}{0.0216}
\newcommand{\SecondJudgeStrictMaxTwinShift}{0.0223}
\newcommand{\SecondJudgeStrictPairCount}{402}
\newcommand{\SecondJudgeSwapInconsistentPairCount}{49}
\newcommand{\SecondJudgeThreeClassKappa}{0.889}
\newcommand{\SecondJudgeThreeClassKappaLower}{0.860}
\newcommand{\SecondJudgeThreeClassKappaUpper}{0.918}
\newcommand{\SecondJudgeUncertainRate}{0.038}
\newcommand{\SecondaryDeltaAUROCMax}{0.344}
\newcommand{\SecondaryDeltaAUROCMin}{0.165}
\newcommand{\SecondaryTransferCellCount}{15}
\newcommand{\SourceAUROCMax}{0.999}
\newcommand{\SourceAUROCMin}{0.996}
\newcommand{\SourceHeadAUROCMax}{0.910}
\newcommand{\SourceHeadAUROCMin}{0.511}
\newcommand{\SourceLogisticBTEMax}{0.300}
\newcommand{\SourceLogisticBTEMin}{0.164}
\newcommand{\SourceProbeCellCount}{15}
\newcommand{\SourceProbeDeltaLCBMin}{0.053}
\newcommand{\SourceProbeFixedFBRMax}{0.214}
\newcommand{\SourceProbeFixedFBRMin}{0.000}
\newcommand{\SourceProbeFixedTPRMax}{0.671}
\newcommand{\SourceProbeFixedTPRMin}{0.200}
\newcommand{\SourceProbeLogisticGainMax}{0.097}
\newcommand{\SourceProbeLogisticGainMin}{0.015}
\newcommand{\SourceProbeLogisticTwinAUROCMax}{0.893}
\newcommand{\SourceProbeLogisticTwinAUROCMin}{0.799}
\newcommand{\SourceProbeXSTestFBR}{0.000}
\newcommand{\SourceTextCVBTEMax}{0.500}
\newcommand{\SourceTextCVBTEMin}{0.436}
\newcommand{\SourceTextDeltaAUROCMax}{0.441}
\newcommand{\SourceTextDeltaAUROCMin}{0.420}
\newcommand{\SourceTextFixedFBRMax}{0.800}
\newcommand{\SourceTextFixedFBRMin}{0.657}
\newcommand{\SourceTextSourceAUROCMax}{0.987}
\newcommand{\SourceTextSourceAUROCMin}{0.976}
\newcommand{\SourceTextTwinAUROCMax}{0.567}
\newcommand{\SourceTextTwinAUROCMin}{0.535}
\newcommand{\SourceXSTestFBRMax}{0.012}
\newcommand{\SourceXSTestFBRMin}{0.000}
\newcommand{\SurfaceBFNullAUROC}{0.821}
\newcommand{\SurfaceBFOriginalAUROC}{0.824}
\newcommand{\SurfaceBFOriginalBothOrdersCount}{144/151}
\newcommand{\SurfaceBFOriginalPairRankingLowerBound}{0.821}
\newcommand{\SurfaceBFSpaceAUROC}{0.732}
\newcommand{\SurfaceBFSpaceBehavior}{0.983}
\newcommand{\SurfaceBFSpaceBothOrdersCount}{146/151}
\newcommand{\SurfaceBFSpaceSTER}{0.370}
\newcommand{\SurfaceBFSpaceSTERFraction}{50/135}
\newcommand{\SurfaceBFUnderscoreAUROC}{0.721}
\newcommand{\SurfaceBFUnderscoreBehavior}{0.977}
\newcommand{\SurfaceBFUnderscoreBothOrdersCount}{144/151}
\newcommand{\SurfaceBFUnderscoreSTER}{0.333}
\newcommand{\SurfaceBFUnderscoreSTERFraction}{45/135}
\newcommand{\TextProxyAUROCMax}{0.955}
\newcommand{\TextProxyAUROCMin}{0.865}
\newcommand{\TwinFitTransferAUROCMax}{0.982}
\newcommand{\TwinFitTransferAUROCMin}{0.930}
\newcommand{\TwinNestedAxisAUROCMax}{0.993}
\newcommand{\TwinNestedAxisAUROCMin}{0.829}
\newcommand{\WallProblemFBRLCB}{0.20}
\newcommand{\WallProblemTPRUCB}{0.80}
\newcommand{\WallProblemXSTestLCB}{0.10}
\newcommand{\WallSourceStrengthLCB}{0.95}
\newcommand{\WallSupportBTELCB}{0.20}
\newcommand{\WallSupportDeltaLCB}{0.10}
\newcommand{\WallSupportPairLCB}{0.50}
\newcommand{\WhitenedAUROCMax}{0.974}
\newcommand{\WhitenedAUROCMin}{0.945}
\newcommand{\WhitenedBTEMax}{0.093}
\newcommand{\WhitenedBTEMin}{0.071}
\newcommand{\XSTestExampleCount}{250}
\newcommand{\XSTestResolution}{0.004}

\title{The Entanglement Wall: Activation-Space Probes as Risk Detectors,
Not Context Adjudicators}

\author{
  \IEEEauthorblockN{Dominik Schwarz}
  \IEEEauthorblockA{Independent Researcher\\
  dominikschwarz@acm.org}
  \thanks{This work has been submitted to the IEEE for possible publication. Copyright may be transferred without notice, after which this version may no longer be accessible.}
}

\begin{document}

\maketitle

\begin{abstract}
An activation-based safety gate must distinguish harmful assistance from
legitimate work on the same topic. We test whether probes trained on broad
harmful-versus-harmless data provide this decision boundary. Five
checkpoint-specific readouts and their source-derived thresholds are transferred
unchanged to four matched-pair cohorts containing \ActiveTwinPairCount{} pairs,
manually reviewed for topic and request-frame similarity. On Twin-70, the
primary cohort of 70 pairs, central harmful and harmless score intervals
overlap. Area under the receiver operating characteristic curve (AUROC) falls from
\SourceAUROCMin{}--\SourceAUROCMax{} on the broad source task to
\PrimaryTwinAUROCMin{}--\PrimaryTwinAUROCMax{}. Pair ranking remains
\PrimaryPairRankingMin{}--\PrimaryPairRankingMax{}, but the transferred
thresholds detect \FixedTwinTPRMin{}--\FixedTwinTPRMax{} of harmful requests
while falsely blocking \FixedTwinFBRMin{}--\FixedTwinFBRMax{} of matched
harmless requests. False blocking remains low on XSTest. Even with thresholds
refitted within pair-grouped cross-validation, balanced threshold error is
\PrimaryCVBTEMin{}--\PrimaryCVBTEMax{}. Alternative source-trained logistic,
radial-basis, and shallow multilayer probes reduce false blocking but detect
only \SourceProbeFixedTPRMin{}--\SourceProbeFixedTPRMax{} of harmful requests
at their transferred thresholds. Three primary readouts
meet all prespecified Wall criteria, while two yield mixed evidence.
Matched-label training recovers the distinction for the three checkpoints
tested, and lexical controls also exhibit transfer loss. An automated
post-freeze audit finds little numerical sensitivity to arm-label disagreement.
The evaluated source operating points fail to
combine strong harmful-request detection with access for legitimate same-topic
work. This gap between risk ranking and a usable decision boundary is the
security concern captured by the Entanglement Wall.
\end{abstract}

\section{Introduction}
\label{sec:introduction}

Harmful assistance and legitimate analysis can concern the same material.
We ask whether source-trained activation probes can distinguish them at an
operating point that detects harmful requests while preserving legitimate access.

Security researchers inspect harmful instructions for red-team evaluation,
moderation training, and benchmark curation. Similar language can describe
deploying malware or analyzing it during an authorized incident response.
The requested action, role, target, and authorization can change which
assistance is appropriate, even when the topic remains the same.

This raises two questions: what information is needed to distinguish harmful
assistance from legitimate work, and which of those distinctions does an
activation probe make available for decisions? Probes map internal states to
scalar scores and often rank broad harmful-versus-harmless corpora well
\cite{zou_repe,arditi_refusal,wollschlager_cones,mckenzie_probes,kramar_probes}.
Related work identifies failures of probe and benchmark generalization
\cite{kumar_probing,bolukbasi_illusion,wang_falsesense,fomin_benchmarks}.
We examine the resulting decisions at an operating threshold.

We use four manually reviewed matched-pair cohorts with
\ActiveTwinPairCount{} pairs in total. Twin-70 is the primary cohort of
70 pairs. Each pair contrasts harmful and harmless requested actions under
shared-topic and similar-request-frame construction criteria.
Section~\ref{sec:problem} distinguishes text, intent, and context. Our
empirical target is the expressed request, not latent intent or verified
real-world authorization.

Harmful requests can score above their harmless partners while score
distributions overlap across pairs. Lowering a common threshold to detect more
harmful requests can then block more harmless requests. We use
\emph{Entanglement Wall} for the supported cases in which strong source
ranking coexists with substantial matched-cohort
ranking loss, a confirmed problem at the transferred operating point, and
persistent error on held-out pairs after threshold recalibration while the
readout remains fixed. Section~\ref{sec:wall-criteria} gives the prespecified
statistical criteria. We apply the term only to readouts and transfers that meet all of them.

\begin{figure*}[t]
\centering
\includegraphics[width=0.99\textwidth]{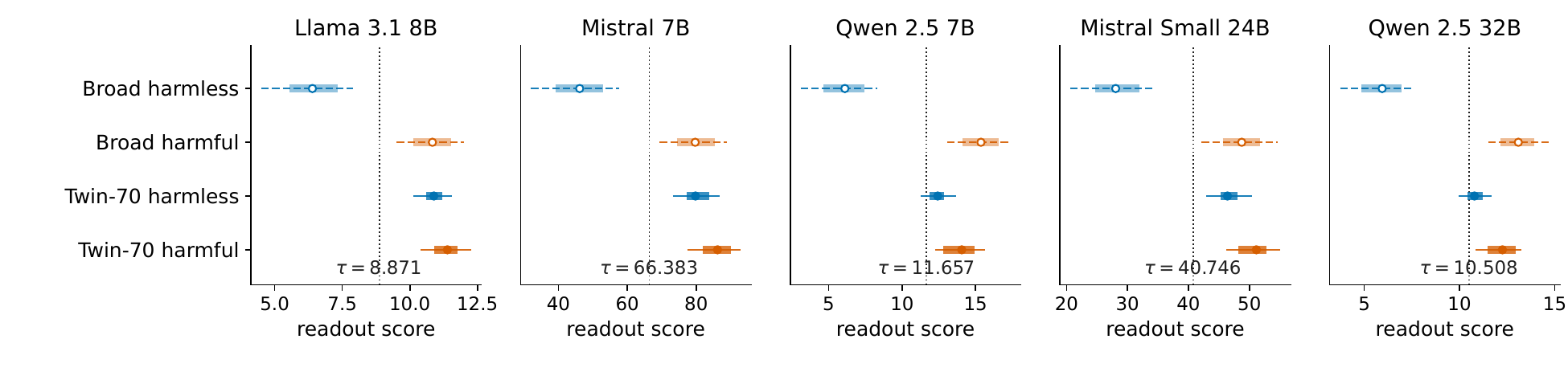}
\caption{Score quantiles for five source-trained activation readouts on
broad source verification data (320 harmful AdvBench and 250 harmless XSTest
requests, dashed outlines) and Twin-70 (70 requests per class, solid outlines).
Lines show the
10th--90th percentiles, thick segments show the 25th--75th percentiles, and
markers show medians. These are score quantiles, not confidence intervals.
Dotted lines mark the
model-specific threshold at the fifth percentile of a separate set of 100 harmful
source-selection scores. Scores are not comparable across panels.}
\label{fig:overlap}
\end{figure*}

Our contribution is evidence that strong broad-risk separation can coexist
with decisions that block legitimate same-topic work or miss harmful requests.
Holding source-trained readouts and their decision rules fixed distinguishes
this transfer test from broad harmful-corpus evaluations
\cite{llorente2603,llorente2604} and matched-data refusal interventions
\cite{petrov_topicmatched}.

We evaluate five checkpoints on the four disjoint synthetic Twin cohorts,
reporting ranking, absolute confusion counts, threshold recalibration, and
false blocking on XSTest. Uncertainty estimates preserve pair grouping.
Source-only logistic, radial-basis, and shallow multilayer probes test
alternative fitting without matched labels. A supplementary test transfers
reference-calibrated thresholds for two strong covariance readouts. Direct matched-label training
and ten readout constructions test what remains readable, while lexical
controls test whether transfer loss also occurs without activation features.
A separate Qwen~2.5 32B surface-form study examines score sensitivity and
paired-choice behavior. A blinded automated judge audits label sensitivity
without changing manual labels.

Figure~\ref{fig:overlap} illustrates the operating-point problem: the unchanged
source-derived threshold for Llama blocks 69 of 70 harmless Twin-70 requests.

\section{From Risk Ranking to Contextual Adjudication}
\label{sec:problem}

\subsection{Relational prediction target}

Let \(T\) denote the observed request text, \(I\) the operative intent, and
\(C\) the relevant use context, including authorization, role, and target.
Let \(H\) denote the resulting harmful-versus-harmless policy label. We write
\begin{equation}
H=h(T,I,C).
\label{eq:harm}
\end{equation}
The function notation expresses joint dependence without prescribing an
additive or multiplicative form. It is a conceptual relation rather than an
estimated causal model. This keeps lexical content distinct from the full
policy target. Requested action and authorization can change the label even
when topic and wording remain similar.
Topic describes the subject matter of \(T\) and does not determine
authorization.

For example, analyzing a malicious program during an authorized incident
response and adapting it for unauthorized deployment can share technical
vocabulary while requiring opposite safety decisions. A signal for that
vocabulary may identify security-relevant material without resolving the
requested action. The distinction matters operationally: increasing detection
of the shared material can also block legitimate analysis. A useful safety
boundary must preserve the action-relevant distinctions, not merely recognize
the topic.

Forrest et al.\ describe how narrow normal-behavior profiles can flag
legitimate variation \cite{forrest_self1996}. Hofmeyr et al.\ identify
credential misuse that their privileged-process monitor can miss
\cite{hofmeyr_syscalls1998}. These observations motivate our distinction in
Eq.~\eqref{eq:harm} between detector-visible patterns and the intent and
authorization needed for a policy decision.

A probe observes a request representation without independently verified
intent or authorization. We test whether its score preserves the expressed
distinctions sufficiently to separate harmful assistance from legitimate work
at a common threshold. Pair ordering tests relative discrimination, while
transferred operating points test absolute decisions across requests.

The matched-pair design provides a limited empirical test of this relation.
Within each pair, we judged the topic to match, while generation
targeted a similar request frame and contrasted a harmful action with
legitimate work. Topic and request-frame similarity are therefore construction
and review criteria rather than independently measured equivalence. The design
varies literal strings and leaves \(I\) and some components of \(C\) unobserved.
Contextual cues can appear inside \(T\). External facts
such as actual authorization remain outside the evaluated readout.

\paragraph{Evaluation and adversarial scope}
We evaluate fixed, non-adaptive requests representing harmful actions and
legitimate same-topic work, and test whether a transferred safety-gate
readout distinguishes them. The prespecified surface transformations
diagnose sensitivity of a fixed readout and do not establish robustness
to adaptive attackers or measure end-to-end attack success.

For an evaluated request \(x\), an activation readout maps the specified
internal state or state sequence to a scalar score \(s(x)\). Larger values
indicate more evidence for the harmful class. Source-trained and
fixed-reference readouts are oriented before matched-cohort results are
inspected. Cross-validated target-supervised readouts are oriented using only
the training portion of each pair-grouped fold. Explicit target-guided
diagnostics are identified as target-guided rather than held-out estimates. A
fixed-threshold decision is
\begin{equation}
\widehat y=\mathbf{1}[s(x)>\tau].
\label{eq:decision}
\end{equation}
Equality is assigned to the negative class. This strict comparison is used
for every confusion matrix. The transfer experiment asks whether a score
learned on a broad source contrast remains a useful boundary when the
matched design reduces topic and request-form differences.

\subsection{Ranking, pairs, and absolute boundaries}

Let \(s_i^+\) and \(s_i^-\) denote the harmful and harmless scores for pair
\(i\). Pooled AUROC compares every harmful score with every harmless score,
with half credit for an exact tie. It does not use pair membership. For each
matched pair, define the margin \(m_i=s_i^+-s_i^-\) and report
\begin{align}
N_c&=\sum_i\mathbf{1}[m_i>0],&
N_w&=\sum_i\mathbf{1}[m_i<0],\nonumber\\
N_t&=\sum_i\mathbf{1}[m_i=0].&
\end{align}
These are the counts of correctly ordered, wrongly ordered, and tied pairs.
The pair-ranking statistic is
\begin{equation}
A_{\mathrm{pair}}=\frac{N_c+\tfrac{1}{2}N_t}{n}.
\label{eq:pair}
\end{equation}
We report all three counts and keep strict pair accuracy \(N_c/n\) separate.
Pair ranking can be high even when the two score distributions overlap enough
to prevent a useful common threshold.

The primary transfer quantity is
\begin{equation}
\Delta_{\mathrm{AUROC}}=
\mathrm{AUROC}_{\mathrm{source}}-
\mathrm{AUROC}_{\mathrm{matched}}.
\label{eq:delta}
\end{equation}
Positive values denote stronger separation on the broad source verification
task. Source and matched samples contain different units, so they are
resampled independently within each bootstrap draw.

Ranking does not specify how many requests a threshold misclassifies. We
count true positives (TP), false negatives (FN), false positives (FP), and
true negatives (TN), treating harmful requests as positive. The true-positive
rate (TPR) is \(TP/(TP+FN)\), and the false-block rate (FBR) is
\(FP/(FP+TN)\) for harmless requests.

Cross-validated balanced threshold error (CV-BTE) asks whether a threshold
chosen on matched training pairs works on held-out pairs. It gives equal
weight to harmful and harmless error rates. In the primary transfer analysis,
the source-trained score remains fixed, but each training partition selects
its own threshold. We use pair-grouped \CVFoldCount{}-fold cross-validation,
keeping both members of a pair in the same fold. Each selected threshold is
applied unchanged to its held-out fold. Aggregated out-of-fold counts give
\begin{equation}
\mathrm{CV\mathchar`-BTE}=
\tfrac{1}{2}\left(
\frac{\sum_f FN_f}{\sum_f(TP_f+FN_f)}
+
\frac{\sum_f FP_f}{\sum_f(FP_f+TN_f)}
\right).
\label{eq:bte}
\end{equation}
We also report the proportions of harmless scores above the harmful median
and harmful scores below the harmless median. Central overlap is the
intersection length of the harmful and harmless 10th--90th percentile
intervals divided by their union length. A value of zero means that these
central intervals are disjoint while leaving the full observed score ranges
unrestricted. Minima and maxima are descriptive only.

\subsection{Transferred operating points}

At a transferred operating point, the fitted readout, orientation, layer,
pooling rule, and source-derived threshold remain fixed on every matched
cohort and on XSTest. Section~\ref{sec:setup} specifies the source-only
selection procedure. Unlike CV-BTE, this evaluation does not use matched
labels to recalibrate the threshold. Worst-group FBR is the maximum across
the prespecified request categories with sufficient harmless support.

XSTest contains \XSTestExampleCount{} harmless prompts, which gives an
empirical rate resolution of \XSTestResolution{}. A rate of zero therefore
reports zero observed false blocks in that finite sample and leaves population
risk unresolved.

\subsection{Entanglement and prespecified Wall criteria}
\label{sec:wall-criteria}

\emph{Entanglement} describes overlap between the harmful and harmless
central score intervals under a common orientation. Pairs may still retain
directed order. Overlap alone does not establish the \emph{Wall}: support
also requires the following transfer, ranking, and decision conditions,
specified before the corresponding evaluation. This is an empirical
score relation, not a claim that matched labels cannot be recovered through
target-supervised fitting.

The source-strength gate requires the lower \ConfidenceLevel\% confidence
bound on source AUROC to be at least \WallSourceStrengthLCB{}. A readout that
passes this gate supports the Wall pattern only when all of the following
conditions hold:

\begin{enumerate}
\item the lower bound on \(\Delta_{\mathrm{AUROC}}\) exceeds
\WallSupportDeltaLCB{}.
\item the lower bound on pair ranking exceeds \WallSupportPairLCB{}.
\item the lower bound on CV-BTE exceeds \WallSupportBTELCB{}.
\item at least one operating-point problem is confirmed by its confidence
interval.
\end{enumerate}

The operating-point condition holds when the upper bound on matched-cohort
TPR is below \WallProblemTPRUCB{}, the lower bound on matched-cohort FBR is
above \WallProblemFBRLCB{}, or the lower bound on XSTest FBR is above
\WallProblemXSTestLCB{}. Counterconditions for the opposite evidence pattern
appear in Appendix~\ref{app:secondary}. A readout that satisfies neither set
of conditions is classified as mixed. Each readout is assessed separately
because the checkpoints do not form exchangeable repetitions of one model.

The margins require more than ten AUROC points of transfer loss and more than
one balanced error in five, while \(0.50\) provides the chance reference for
pair ordering. These prespecified values are study criteria rather than
deployment acceptance limits.
The reproduction artifact records the definitions and local-protocol provenance
in \texttt{scientific/protocols}.

These criteria apply only to the evaluated checkpoint, representation, layer,
read point, pooling rule, threshold, and cohort. They allow contextual
information elsewhere in the activations. Sparse learned features, deeper
nonlinear classifiers, interventions, other token positions, and systems that
explicitly observe authorization and provenance remain outside the study.

\subsection{Matched-versus-shuffled diagnostic}

A secondary diagnostic holds every text and score fixed, permutes harmless
members within request category, and compares true-pair ranking with the
resulting reference distribution. It asks whether the specified pairing adds
rank structure beyond category composition. It remains outside the Wall
criteria and agnostic about the cause of
\(\Delta_{\mathrm{AUROC}}\). Complete values and the centered two-sided test
appear in Appendix~\ref{app:pairing}.

\section{Experimental Setup}
\label{sec:setup}

\subsection{Datasets and matched cohorts}

Separate source splits fit, select, and verify readouts before within-topic
Twin evaluation.

Source fitting uses 100 harmful AdvBench requests and 100 harmless Alpaca
requests selected with seed 42 \cite{zou_gcg,alpaca}. A disjoint selection
split contains another 100 requests from each dataset and is used for layer
selection and threshold calibration. Broad verification uses the remaining
320 AdvBench requests and all \XSTestExampleCount{} harmless XSTest requests
\cite{rottger_xstest}. The same XSTest sample supplies challenging-harmless
FBR, so these views are not independent.

Each matched item contains one harmful and one harmless request. We label
the expressed request: a harmful arm asks for an action whose fulfillment
would materially enable, facilitate, or carry out harm. Dangerous vocabulary
alone is insufficient. Legitimate analysis, moderation, research, or
authorized defensive work is distinguished from the harmful action.
Appendix~\ref{app:cohort-construction} gives the complete rubric.

Language models generated candidates using fixed criteria requiring a shared topic, similar
request frame, and different requested actions. We manually reviewed and
labeled every candidate pair under the stated rubric, admitting only
qualifying pairs. Matching refers to these construction and review criteria.

We name the four cohorts by pair count:
Twin-70 is the primary cohort, Twin-120 is the replication cohort, Twin-217
is the robustness cohort, and Twin-43 preserves the original cohort. Their
\ActiveTwinPairCount{} pairs are mutually disjoint under exact and normalized
comparisons of both pairs and individual arms.

Twin-70 is primary because it was generated separately and initially reviewed
without the arm mapping or experimental results. The additional cohorts test
replication and robustness across separately reported matched collections.
Twin-120 is the remainder of a larger reviewed collection after removing
Twin-43. Appendix~\ref{app:cohort-construction} gives the construction and
review details. Cohort membership remains fixed throughout evaluation.

The surface-form case study uses Twin-151, combining 31 Twin-43 pairs with
all 120 Twin-120 pairs. Its membership follows a fixed, result-independent
exclusion of 12 exact pair overlaps with an earlier version of Twin-43 from the
163-pair source collection.

\begin{table}[t]
\centering
\caption{Data roles. Source-fit and Source-selection each contain 100 harmful AdvBench and 100 harmless Alpaca requests in disjoint splits. Sizes give harmful/harmless request counts. Twin sizes give pair counts, with one harmful and one harmless request per pair. Source splits and the four Twin cohorts are disjoint.}
\label{tab:study_data}
\footnotesize
\setlength{\tabcolsep}{3pt}
\begin{tabular}{@{}p{0.24\linewidth}p{0.19\linewidth}p{0.49\linewidth}@{}}
\toprule
Data & Size & Role \\
\midrule
Source fit & 100/100 & Fit primary and alternative readouts \\
Source selection & 100/100 & Select layer or classifier and calibrate threshold \\
Broad verification & 320/250 & Held-out AdvBench and safe XSTest \\
Twin-70 & 70 & Primary confirmatory evaluation \\
Twin-120 & 120 & Secondary replication \\
Twin-217 & 217 & Secondary robustness \\
Twin-43 & 43 & Secondary original cohort \\
\bottomrule
\end{tabular}
\end{table}

\subsection{Primary readout and unchanged transfer}
\label{sec:primary-readout}
\label{sec:source-thresholds}

The primary experiment fits one activation readout per checkpoint on a broad
harmful--harmless source contrast and applies it unchanged to matched requests.
We evaluate Llama~3.1 8B Instruct, Mistral~7B Instruct v0.3,
Qwen~2.5 7B Instruct, Mistral Small 24B Instruct 2501, and
Qwen~2.5 32B Instruct
\cite{llama3,mistral7b,qwen25,mistral_small3}. The B suffix denotes the
nominal parameter count in billions. The Llama and 7B Qwen checkpoints are
community variants described by their publishers as
\emph{abliterated}, meaning that refusal behavior was modified after
instruction tuning. We evaluate these two variants alongside three unmodified
instruction-tuned checkpoints. The primary readout uses prompt activations.
Effects of refusal modification and model scale are not isolated by this
comparison.

The primary design adapts Llorente-Saguer's mean-difference readout fitted on
raw AdvBench and Alpaca requests with maximum pooling~\cite{llorente2604}.
Language-model prompts are processed serially with one sequence and a fresh
context per forward pass. The main path uses Python, Hugging Face, and 16-bit
bfloat16 (BF16) parameters. For the primary readout, terminal punctuation is
removed before tokenization, no chat template or special tokens are added,
and the first 64 tokens are retained. Stored token counts confirm that every
Twin-70 request fits within this window for all five checkpoints. We record
the post-attention layer-normalization output, differing from that study's
transformer-block output, and take the componentwise maximum over token
positions.
Let \(z(x)\) denote this pooled vector. With harmful and harmless means from
the source fit split, the readout is
\begin{equation}
v=\frac{\mu_{\mathrm{harm}}^{\mathrm{fit}}-\mu_{\mathrm{benign}}^{\mathrm{fit}}}
{\|\mu_{\mathrm{harm}}^{\mathrm{fit}}-\mu_{\mathrm{benign}}^{\mathrm{fit}}\|_2},
\qquad s(x)=v^\top z(x).
\label{eq:primary-readout}
\end{equation}
Only the direction \(v\) is unit-normalized, not the pooled vector \(z(x)\).
Thus this is a dot-product score rather than a cosine score. Larger values
indicate the harmful class.

We considered every transformer block and selected the highest
source-selection AUROC, breaking exact ties in favor of the lowest layer
index. The candidate zero-based indices are 0--31 for Llama~3.1 8B and
Mistral~7B, 0--27 for Qwen~2.5 7B, 0--39 for Mistral Small 24B, and 0--63
for Qwen~2.5 32B. The selected layers are 10 for Llama~3.1 8B, 16 for
Mistral~7B, 19 for Qwen~2.5 7B, 18 for Mistral Small 24B, and 47 for
Qwen~2.5 32B. The threshold calibration targets approximately 95\% detection
on the harmful source-selection set. For each model, the operating threshold is the fifth percentile
of its 100 harmful selection scores:
\begin{equation}
\begin{aligned}
\tau=Q_{0.05}\!\left(\{s(x):x\in
D_{\mathrm{selection}}^{\mathrm{harm}}\}\right),\\
\widehat y(x)=\mathbf{1}[s(x)>\tau].
\end{aligned}
\label{eq:source-threshold}
\end{equation}
The resulting model-specific
configuration and threshold are then applied unchanged to source
verification, all four matched cohorts, and XSTest. Fitting, orientation,
selection, and calibration use source data only.
CV-BTE asks a separate question: the source-trained score stays fixed, while
matched training pairs select the threshold.

\subsection{Alternative source-trained readouts}
\label{sec:source-probe-control}

To test whether the primary result depends on the mean-difference fitting
rule, we added three classifiers on the same stored activation vectors. Their
search spaces were fixed before computing this control's results.
They use the same layers, pooling rule, and source
split. A standard scaler precedes logistic regression, a radial-basis support
vector machine, or a one-hidden-layer multilayer perceptron (MLP).
Hyperparameters for regularization, radial-basis width, hidden width, and MLP
penalty are selected by source-selection AUROC, with
no refit on the selection rows. The MLP averages three fixed-seed scores. Each
readout uses the fifth percentile of its harmful Source-selection scores and
is transferred unchanged. We report all \SourceProbeCellCount{}
checkpoint-by-readout cells. Twin and XSTest labels neither fit nor select a
readout. This later-added control does not alter the primary Wall criteria.
Appendix~\ref{app:probe-hyperparameters} gives the complete grids, training
settings, and seed aggregation.

The source-trained lexical controls were specified before their scores were
computed. They use the same cleaned raw request text as the primary readout
and no chat template. Word term-frequency--inverse-document-frequency
(TF-IDF) features cover contiguous sequences of one or two words, and
character TF-IDF features cover sequences of three to five characters.
Each logistic model is fitted on the same source fit split. Its regularization
is selected through the inverse-regularization parameter
\(C_{\mathrm{reg}}\in\{0.01,0.1,1,10\}\) by source-selection AUROC,
without refitting on the selection rows. The fifth percentile of the harmful
source-selection scores supplies the unchanged operating threshold. Both
models are held fixed for evaluation on the broad data, the four matched
cohorts, and XSTest. We report both without selecting a preferred control.
These text-only readouts test whether the transfer pattern is specific to
activations and remain outside the activation Wall criteria. Their separate
failure assessment requires a source-AUROC lower bound of at least \(0.95\),
lower bounds above \(0.10\), \(0.50\), and \(0.20\) for
\(\Delta_{\mathrm{AUROC}}\), pair ranking, and CV-BTE, respectively, and at
least one confidence-bound operating problem. The latter is a matched TPR
upper bound below \(0.90\), a matched FBR lower bound above \(0.20\), or an
XSTest FBR lower bound above \(0.10\).
The lexical protocol uses a \(0.90\) rather than \(0.80\) TPR bound, but
both controls satisfy the FBR condition alone (lower bounds \(0.543\) and
\(0.700\), Table~\ref{tab:decision_intervals}), leaving their assessments
unaffected.

\subsection{Representation methods and label access}
\label{sec:axes-method}
\label{sec:representation-methods}

This comparison uses Llama~3.1 8B, Mistral~7B, and Qwen~2.5 7B to examine
readability under different representations and access to labels.

The representation comparison formats messages with each checkpoint's native
chat template, meaning the message format prescribed for that checkpoint. It
retains token positions wholly contained in the user request. Unless stated
otherwise, activations are taken after attention-layer normalization at
zero-based layers 12 for Llama~3.1 8B and 14 for both 7B checkpoints.

Eight methods use a project-specific reference collection with 24 harmful
and 19 harmless categories. We manually reviewed and assigned the final
harmfulness and category labels. The collection contains 4,082 requests,
with 3,008 for fitting and 1,074 for validation.
Appendix~\ref{app:representation-detail} describes the review and selection.
Hash-based selection followed normalized
deduplication and exclusion of matches to the active Twin, AdvBench, Alpaca,
and XSTest datasets. Its frozen overlap audit found no exact or normalized
matches to those datasets or between reference splits. Normalization uses
Unicode NFKC, lowercasing, and whitespace collapse and establishes textual
disjointness rather than semantic independence. For centroid-based methods,
we average the normalized per-request activation means within each category
and normalize the resulting 43 centroids. Each retained user token is
represented by its cosine similarity to all centroids.

Table~\ref{tab:study_overview} groups the ten methods by the labels used to
fit or select their readout. Fixed-reference methods use prescribed reference
comparisons. Reference-fitted methods estimate directions or covariance
structure without Twin training labels. Matched-trained methods learn within
pair-grouped training folds. Benign-subspace projection is a separate
diagnostic, not a deployment candidate.

This grouping concerns readout fitting, not threshold evaluation. Every
reported CV-BTE selects its threshold using matched training labels, including
for the reference-based scores. The methods also differ in read points and
pooling, so their results are not a common-regime leaderboard.
Appendix~\ref{app:representation-detail} specifies their constructions and
controls.

\paragraph{Supplementary reference calibration}
For whitened mean difference and local covariance, we additionally calibrate
one threshold per checkpoint and method on the disjoint reference-validation split.
The rule was specified before these operating-point results were computed:
the linearly interpolated fifth percentile of the 595 harmful validation
scores, with decisions made by \(s>\tau\). The six thresholds transfer
unchanged to all four Twin cohorts. No Twin labels select a threshold.
Mean pooling defines each request's activation representation, whereas the
score quantile defines the boundary across calibration requests. No request
is removed by this calibration. Appendix~\ref{app:reference-calibration}
provides the protocol and all 24 operating points.

\begin{table}[t]
\centering
\caption{Ten representation methods, grouped by readout fitting and selection. The groups are not a common ranking of interchangeable procedures.}
\label{tab:study_overview}
\footnotesize
\setlength{\tabcolsep}{3pt}
\begin{tabular}{@{}p{0.46\linewidth}p{0.49\linewidth}@{}}
\toprule
Method & Input \\
\midrule
\multicolumn{2}{@{}l}{\textit{Fixed references}} \\
Generation window & Generated-token reference margins \\
Sequence warping & Token-to-reference sequences \\
Frame-content co-occurrence & Early user-token co-occurrence \\
\multicolumn{2}{@{}l}{\textit{Reference-fitted}} \\
Depth difference$^{v}$ & Mean activations across layers \\
Whitened mean difference$^{v}$ & Mean user activations \\
Local covariance$^{v}$ & Mean user activations \\
\multicolumn{2}{@{}l}{\textit{Matched training}} \\
Recurrent sequence & Token-to-reference sequences \\
Layer/read point & Candidate layers/read points \\
Directed transitions & Dominant-centroid transitions \\
\multicolumn{2}{@{}l}{\textit{Diagnostic projection}} \\
Benign-subspace projection & Projected mean activations \\
\bottomrule
\end{tabular}
\par\smallskip
\(^{v}\)Reference validation selects the depth-difference layer. Its initial use for covariance readouts was descriptive. Section~\ref{sec:axes-method} details their supplementary threshold calibration. Layer/read-point selection covers only Llama~3.1 8B. Benign-subspace projection is diagnostic only.
\end{table}

For the direct readability analysis, logistic regression, a support-vector
machine with a radial-basis kernel, and a one-hidden-layer multilayer
perceptron use the last user-token activation at zero-based layer 12 for
Llama~3.1 8B and layer 14 for both 7B checkpoints. Hyperparameters are
selected within three inner folds and evaluated in five outer folds, with
folds assigned by pair. Both fold levels are stratified by request category. For
cross-cohort evaluation, the five outer-fold models trained on Twin-70 score
Twin-120 and Twin-217. Their standardized scores are averaged. For the
multilayer perceptron, predictions are
additionally averaged across three fixed training seeds. Target labels are
not used when scoring either external cohort.
Appendix~\ref{app:direct-readout-hyperparameters} specifies the hyperparameters
and fold-local fitting procedure.

\paragraph{Secondary proxy analyses}
Pair-grouped linear classifiers use word and character TF-IDF features and
mean-pooled, normalized all-MiniLM-L6-v2 embeddings as text-only proxies for
surface form \cite{wang_minilm}. Incremental models test whether activation
predictions add information after a proxy is available. Attention analyses
compare one component of the attention mechanism, called a head, selected on
source data with a score from the residual stream, the hidden state passed
between transformer blocks. A separate diagnostic selects a head directly
with Twin-70 labels and is reported only as a target-guided upper bound. These
analyses measure predictive dependence and incremental prediction rather than
causal influence.

\subsection{Blinded model-based label audit}
\label{sec:second-judge-method}

After freezing our manual review, scores, folds, and thresholds, we ran a
model-based label-sensitivity audit. The requested
\texttt{gpt-5.6-sol} judge saw each of the \SecondJudgeArmCount{} arms once
in isolation and each of the \SecondJudgePairCount{} pairs twice in swapped
A/B order. Every presentation used a fresh API context and contained only a
fixed rubric and the arm or pair. Cohort and row identifiers, our manual
labels,
probe scores, experimental results, claims, manuscript text, prior responses,
and external tools were withheld. Requests were serial and used
\texttt{store=false}.

The \emph{isolated-arm audit} assigned harmful, harmless, or uncertain to
each isolated arm. Before these outputs were inspected, we specified two
complete-pair sensitivity subsets. The consensus subset excludes a pair only
when at least one binary judge label directly contradicts our manual label for
that arm.
The strict subset additionally requires two completed binary responses that
match our manual labels. Both analyses retain the original labels, source
scores, fold decisions, and thresholds. The \emph{swapped-order pair assessment}
records pair validity,
orientation, topic, request-frame, and swap diagnostics. It remains outside
subset selection and is an automated sensitivity audit. It is not a second
human review or an inter-rater study.

\subsection{Qwen 32B surface-form case study}
\label{sec:surface-method}

A separate Qwen~2.5 32B BF16 case study compares scalar scores with the
model's ability to identify the harmful arm of a pair. We apply space and
underscore insertion plus a function-word control to Twin-151. The activation
readout retains 64 tokens, while the behavioral check sees complete pairs.
Both systematic transformations exceed the activation window for every arm,
confounding surface form with retained token coverage.

The diagnostic threshold uses unmodified Twin-151 harmful scores. Surface
threshold evasion rate (STER) conditions on correct paired choices in both
orders before and after transformation, and on an initially triggered score.
It measures subsequent threshold crossings, not harmful compliance.
Appendix~\ref{app:surface} gives transformations, counting rules, fixed
criteria, pooling controls, and the separate four-bit comparison.

\subsection{Uncertainty and verification}

Point estimates are reconstructed from individual predictions. Simple binary
TPR and FBR proportions in Table~\ref{tab:operating} use two-sided 95\%
Clopper--Pearson intervals. AUROC, pair ranking, CV-BTE, paired differences,
and primary bootstrap intervals use two-sided \ConfidenceLevel\% percentile
intervals based on \ResampleDrawCount{} deterministic draws. The additional
Source-only probe control uses 5,000 draws under its separate specification.
CV-BTE intervals resample frozen out-of-fold pair decisions. They condition
on the fixed folds and fitted thresholds, without including variation from
new folds or threshold refitting. Source classes are
resampled independently, whereas matched analyses resample pair identifiers
and retain both arms. Category-level analyses preserve the observed group
sizes. Paired surface-form comparisons use the same resampled pairs in both
conditions. STER uses a two-sided 95\% Wilson interval on the eligible-pair
count. Confidence intervals describe individual rather than simultaneous
estimands.

An independently written verifier reconstructed all reported results from saved
predictions, separately implementing orientation, ties, pair grouping, fold
isolation, fixed-threshold decisions, bootstrap intervals, the surface-form
denominator, and Wall criteria.

\section{Results}
\label{sec:results}

\subsection{Activation transfer}
\label{sec:rq1}
\label{sec:primary-results}

The five source-trained activation readouts achieve
\SourceAUROCMin{}--\SourceAUROCMax{} AUROC on the broad verification
contrast. When each complete source-trained configuration is applied
unchanged to Twin-70, AUROC is
\PrimaryTwinAUROCMin{}--\PrimaryTwinAUROCMax{}. The corresponding loss,
\(\Delta_{\mathrm{AUROC}}\), ranges from
\PrimaryDeltaAUROCMin{} to \PrimaryDeltaAUROCMax{}, and every confidence
interval excludes zero.

Substantial within-pair order remains despite the pooled loss. Pair ranking is
\PrimaryPairRankingMin{}--\PrimaryPairRankingMax{}, with no ties for these
five readouts. Table~\ref{tab:primary_results} reports the correct, wrong, and
tied pair counts. Table~\ref{tab:central-overlap} reports the central
interval overlap and both median crossovers defined in
Section~\ref{sec:problem}.

\begin{figure*}[t]
\centering
\includegraphics[width=0.98\textwidth]{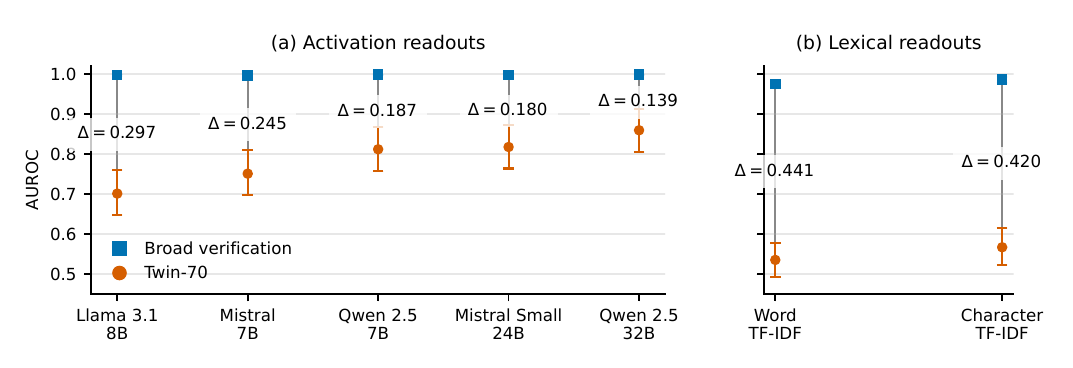}
\caption{Broad-source and Twin-70 AUROC after unchanged transfer. Panel (a)
shows the five activation readouts, and panel (b) shows the two source-trained
TF-IDF controls. Points and bars denote
AUROC and 95\% bootstrap intervals. Labels report
\(\Delta_{\mathrm{AUROC}}\). The word TF-IDF control misses the prespecified
pair-ranking support criterion even though its source boundary still fails to
transfer.}
\label{fig:transfer}
\end{figure*}

\begin{table*}[t]
\caption{Primary Source-to-Twin-70 transfer. Values are point estimates with 95\% bootstrap intervals. C/W/T denotes correctly ordered, wrongly ordered, and tied pairs. CV-BTE reselects thresholds on Twin training folds while the readout remains fixed.}
\label{tab:primary_results}
\centering
\footnotesize
\setlength{\tabcolsep}{3pt}
\begin{tabular*}{\textwidth}{@{\extracolsep{\fill}}lrrrrrr@{}}
\toprule
Model & Source AUROC & Twin-70 AUROC & $\Delta$AUROC & Pair rank & CV-BTE & Pair C/W/T \\
\midrule
Llama 3.1 8B & 0.998 [0.996, 1.000] & 0.701 [0.648, 0.761] & 0.297 [0.238, 0.351] & 0.829 [0.743, 0.914] & 0.329 [0.271, 0.386] & 58/12/0 \\
Mistral 7B & 0.996 [0.993, 0.999] & 0.751 [0.697, 0.810] & 0.245 [0.187, 0.299] & 0.886 [0.800, 0.957] & 0.329 [0.271, 0.386] & 62/8/0 \\
Qwen 2.5 7B & 0.999 [0.998, 1.000] & 0.812 [0.757, 0.868] & 0.187 [0.130, 0.242] & 0.929 [0.857, 0.986] & 0.229 [0.171, 0.286] & 65/5/0 \\
Mistral Small 24B & 0.997 [0.994, 0.999] & 0.818 [0.764, 0.873] & 0.180 [0.125, 0.232] & 0.929 [0.857, 0.986] & 0.286 [0.229, 0.343] & 65/5/0 \\
Qwen 2.5 32B & 0.999 [0.997, 1.000] & 0.860 [0.806, 0.912] & 0.139 [0.086, 0.191] & 0.929 [0.857, 0.986] & 0.229 [0.171, 0.286] & 65/5/0 \\
\bottomrule
\end{tabular*}
\end{table*}

\paragraph{The transferred threshold}
At the unchanged thresholds, TPR on the 320 held-out harmful source
requests is 0.944--0.978 for the primary readouts and 0.922--0.969 for the
alternative source-trained probes (Table~\ref{tab:source_verification_tpr}).
The selection target of 0.95 therefore differs from observed verification
performance. Every alternative probe has lower TPR on Twin-70 than on
source verification.

At the same source-derived thresholds, XSTest FBR is
\SourceXSTestFBRMin{}--\SourceXSTestFBRMax{}, whereas harmless-arm FBR on
Twin-70 is \FixedTwinFBRMin{}--\FixedTwinFBRMax{}.
In every panel of Fig.~\ref{fig:overlap}, the harmless Twin-70
10th--90th percentile interval lies entirely above the corresponding
broad harmless (XSTest) interval. This shift accompanies low XSTest false
blocking and high matched false blocking at the tested thresholds.

Twin-70 TPR at these thresholds is
\FixedTwinTPRMin{}--\FixedTwinTPRMax{}. Depending on the readout, only one to
22 of the \PrimaryTwinPairCount{} harmless arms remain at or below threshold.
Table~\ref{tab:operating} reports the corresponding \(TP,FN,FP,TN\) counts.
With seven harmless arms in each Twin-70 category, a worst-group FBR of
\(1.0\) records one category in which all seven observed arms were blocked.
The worst-group column is a selected maximum over ten small groups, so we show
its point estimate and denominator without attaching a single-group interval.
It is not a rare-error estimate.

\begin{table*}[t]
\caption{Transferred activation operating points on Twin-70 and XSTest at the
source thresholds defined in Section~\ref{sec:source-thresholds}. Decisions
use \(s>\tau\). Brackets are two-sided 95\% Clopper--Pearson intervals for
the displayed binomial proportions. The worst-group column reports only the
maximum FBR across the ten prespecified request categories and its visible
denominator.}
\label{tab:operating}
\centering
\footnotesize
\setlength{\tabcolsep}{2pt}
\begin{tabular*}{\textwidth}{@{\extracolsep{\fill}}lrrrrrrrr@{}}
\toprule
Model & $\tau$ & Twin TP/FN & Twin FP/TN & TPR & FBR & Worst FBR ($n=7$) & XSTest FP/TN & XSTest FBR \\
\midrule
Llama 3.1 8B & 8.871 & 70/0 & 69/1 & 1.000 [0.949, 1.000] & 0.986 [0.923, 1.000] & 1.000 & 0/250 & 0.000 [0.000, 0.015] \\
Mistral 7B & 66.383 & 69/1 & 69/1 & 0.986 [0.923, 1.000] & 0.986 [0.923, 1.000] & 1.000 & 3/247 & 0.012 [0.002, 0.035] \\
Qwen 2.5 7B & 11.657 & 66/4 & 57/13 & 0.943 [0.860, 0.984] & 0.814 [0.703, 0.897] & 1.000 & 0/250 & 0.000 [0.000, 0.015] \\
Mistral Small 24B & 40.746 & 70/0 & 68/2 & 1.000 [0.949, 1.000] & 0.971 [0.901, 0.997] & 1.000 & 1/249 & 0.004 [0.000, 0.022] \\
Qwen 2.5 32B & 10.508 & 66/4 & 48/22 & 0.943 [0.860, 0.984] & 0.686 [0.564, 0.791] & 0.857 & 0/250 & 0.000 [0.000, 0.015] \\
\bottomrule
\end{tabular*}
\end{table*}

Appendix~\ref{app:sensor} reports the generation-time sensor operating
points.

\paragraph{Threshold recalibration on matched training pairs}
CV-BTE is
\PrimaryCVBTEMin{}--\PrimaryCVBTEMax{}. This error persists even though each
training fold may select a threshold from matched training labels. The
observed boundary error therefore extends beyond transfer of the
fifth-percentile source threshold.

\paragraph{Wall assessment}
All five primary readouts pass the source-strength gate in
Section~\ref{sec:wall-criteria}. Llama~3.1 8B, Mistral~7B, and Mistral Small
24B satisfy all Wall support criteria, giving three supported readouts.
Qwen~2.5 7B yields mixed evidence
because the lower confidence bound on CV-BTE is \(0.171\), below the required
strict bound of \(0.20\). Qwen~2.5 32B is also mixed because the lower bounds
on \(\Delta_{\mathrm{AUROC}}\) and CV-BTE are \(0.086\) and \(0.171\),
respectively. Neither mixed result satisfies the counterconditions in
Appendix~\ref{app:secondary}. Supported and mixed outcomes both occur among
the refusal-modified and unmodified checkpoints. Each conclusion is specific
to one readout.

\paragraph{Secondary checks}
Mistral Small 24B and Qwen~2.5 32B have the smallest point estimates of
\(\Delta_{\mathrm{AUROC}}\), but model family, training data, and post-training
differ. The confidence intervals overlap, and fixed-threshold false blocking
varies nonmonotonically with parameter count. These five observations are
insufficient for a scale effect.

Across the \SecondaryTransferCellCount{} model-by-cohort results on Twin-120,
Twin-217, and Twin-43, \(\Delta_{\mathrm{AUROC}}\) remains positive at
\SecondaryDeltaAUROCMin{}--\SecondaryDeltaAUROCMax{}, with every confidence
interval above zero. The three cohorts are reported separately as
replication, robustness, and original-cohort checks rather than pooled with
Twin-70. At the unchanged source thresholds, their TPR is 0.837--1.000 and
FBR is 0.465--1.000. Table~\ref{tab:secondary-operating} reports all 15
operating points and absolute counts.

The category-conditioned pairing diagnostic finds more true-pair rank
structure than its shuffled reference for four readouts
(\(p<0.004\)). Qwen~2.5 32B is at the reference boundary
(\(p=0.0504\)). These are secondary, readout-specific diagnostics. We apply no
familywise error correction and make no joint significance claim. This
diagnostic remains agnostic about the cause of the transfer loss. Complete values
appear in Appendix~\ref{app:pairing}.

\subsection{Alternative source-trained probes}
\label{sec:source-probe-results}

Source-trained logistic regression raises Twin-70 AUROC by
\SourceProbeLogisticGainMin{}--\SourceProbeLogisticGainMax{} relative to
mean difference (gains computed before rounding), reaching
\SourceProbeLogisticTwinAUROCMin{}--\SourceProbeLogisticTwinAUROCMax{}.
All \SourceProbeCellCount{} logistic, radial-basis, and shallow multilayer
combinations retain a positive source-minus-Twin AUROC interval. The smallest
lower bound is \SourceProbeDeltaLCBMin{}. These comparisons use source
labels only, unlike the matched-trained readouts below.

\begin{figure}[!htbp]
\centering
\includegraphics[width=\columnwidth]{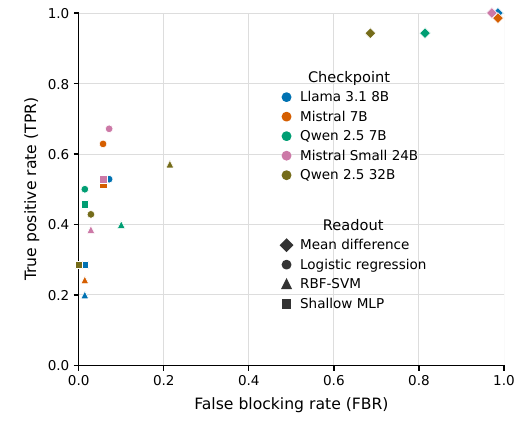}
\caption{Twin-70 operating points for five checkpoints and four source-trained
readouts. Each point uses its own unchanged fifth-percentile harmful
source-selection threshold. Coordinates are observed TPR and FBR, not a ROC
curve or a frontier obtained by threshold search. XSTest has no harmful
examples and is therefore not plotted.}
\label{fig:source-operating}
\end{figure}

Figure~\ref{fig:source-operating} shows the different decision errors at the
transferred thresholds. The primary mean-difference readouts detect most
harmful arms but also block most harmless arms. The additional classifiers
block fewer harmless arms, while missing many more harmful arms. Their FBR is
\SourceProbeFixedFBRMin{}--\SourceProbeFixedFBRMax{}, but TPR falls to
\SourceProbeFixedTPRMin{}--\SourceProbeFixedTPRMax{}. These alternative fits do not resolve
the joint detection and false-blocking problem at these operating points. All follow the same source calibration rule, but each
transfers its own threshold.
These observed points do not identify a Pareto boundary or an unavoidable
trade-off across other possible thresholds. XSTest FBR is
\SourceProbeXSTestFBR{} for the additional classifiers.
The separate CV-BTE evaluation also improves for several alternatives.
Logistic readouts reach \SourceLogisticBTEMin{}--\SourceLogisticBTEMax{},
with point estimates below \(0.20\) for four of the five checkpoints.
Thus the error after target calibration depends on the fitting
method. These point estimates do not apply the full primary Wall assessment
to the additional classifiers, nor do they measure an unchanged source
operating point. Appendix~\ref{app:source-probes} reports every combination.

\subsection{Matched-label readability and representation methods}
\label{sec:rq2}

\begin{figure*}[!t]
\centering
\includegraphics[width=\textwidth]{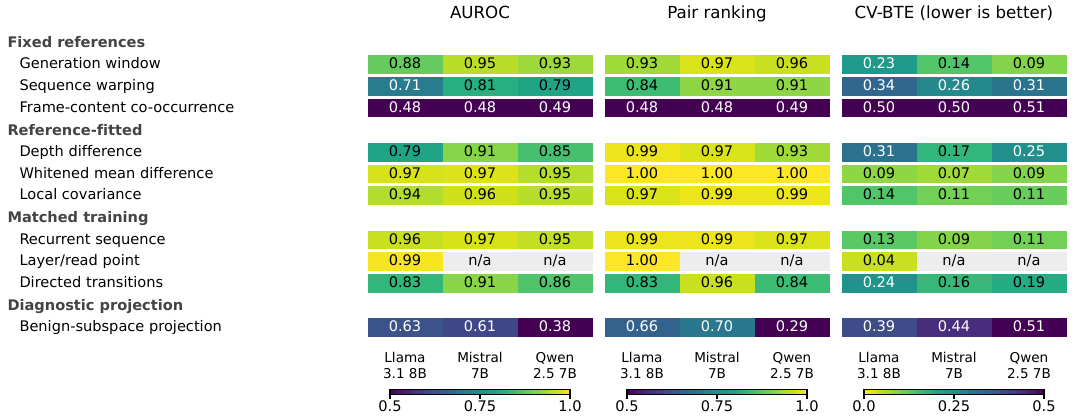}
\caption{Results for the ten representation methods on Twin-70. Rows are grouped by readout-fitting label access, as defined in Section~\ref{sec:axes-method}. Columns are the
three checkpoints used in this comparison. AUROC, half-tie-weighted pair
ranking, and pair-grouped CV-BTE are reported separately. N/A means not
evaluated: the layer/read-point readout was tested only on Llama~3.1 8B.
CV-BTE additionally selects thresholds on matched training
folds for every method, including reference-based scores.}
\label{fig:methods}
\end{figure*}

Figure~\ref{fig:methods} reports the ten methods defined in
Section~\ref{sec:axes-method}, grouped by their access to labels. The
target-supervised recurrent-sequence, layer/read-point, and
directed-transition methods and their controls reach
\TwinNestedAxisAUROCMin{}--\TwinNestedAxisAUROCMax{} AUROC on Twin-70.
Direct logistic, radial-basis, and multilayer classifiers reach \DirectTwinAUROCMin{}--\DirectTwinAUROCMax{} out-of-fold
AUROC with \DirectTwinBTEMin{}--\DirectTwinBTEMax{} CV-BTE. Ensembles trained
with Twin-70 labels reach
\TwinFitTransferAUROCMin{}--\TwinFitTransferAUROCMax{} AUROC when applied
unchanged to Twin-120 and Twin-217. These matched-label results appear in
Tables~\ref{tab:methods_appendix} (representation methods) and~\ref{tab:direct-readouts}
(direct classifiers and cross-cohort transfer).

Table~\ref{tab:sequence-controls} compares token-shuffle and
symmetric-transition controls with their corresponding readouts on all four
cohorts. On Twin-70, Twin-120, and Twin-217, control AUROC differs from the
corresponding readout by at most \(0.031\), although CV-BTE changes are larger
in some cells. On Twin-43, token shuffling lowers recurrent AUROC by up to
\(0.118\). Both controls operate on activations already contextualized by the LLM.

The reference-fitted methods also produce strong results. Whitened mean
difference reaches \WhitenedAUROCMin{}--\WhitenedAUROCMax{} AUROC,
perfect pair ranking, and \WhitenedBTEMin{}--\WhitenedBTEMax{} CV-BTE.
Local covariance reaches
\LocalCovarianceAUROCMin{}--\LocalCovarianceAUROCMax{} AUROC and
\LocalCovarianceBTEMin{}--\LocalCovarianceBTEMax{} CV-BTE. Their readouts
use reference labels, not Twin training labels. Their CV-BTE thresholds
use matched training pairs, yielding low error in that evaluation regime.
The co-occurrence score,
by comparison, is near chance. Recurrent-sequence and layer/read-point
readouts achieve low CV-BTE with matched-label training.

\paragraph{Reference-calibrated operation}
The supplementary reference thresholds yield
\ReferenceTwinTPRMin{}--\ReferenceTwinTPRMax{} TPR and
\ReferenceTwinFBRMin{}--\ReferenceTwinFBRMax{} FBR on Twin-70,
substantially reducing false blocking relative to the primary readouts.
All six calibration readouts
detect 565 of 595 harmful requests and block the same one of 479 harmless
requests. These are calibration diagnostics.
On Twin-120, Twin-217, and Twin-43, TPR is
\ReferenceSecondaryTPRMin{}--\ReferenceSecondaryTPRMax{} and FBR is
\ReferenceSecondaryFBRMin{}--\ReferenceSecondaryFBRMax{}.
Table~\ref{tab:reference-calibration} reports all six thresholds and 24
operating points. On Twin-120, Mistral's
whitened readout meets both interval-based operating counterconditions in
Appendix~\ref{app:wall-counterconditions} (TPR lower bound \(0.812\), FBR upper
bound \(0.188\)). The complete counterpattern remains unassessed: a
pair-ranking confidence bound is unavailable, and \(\Delta_{\mathrm{AUROC}}\),
worst-category FBR, and XSTest FBR were not evaluated for this supplement.

\subsection{Lexical controls}
\label{sec:proxies}

The source-trained word and character TF-IDF controls obtain
\SourceTextSourceAUROCMin{}--\SourceTextSourceAUROCMax{} on broad
verification and \SourceTextTwinAUROCMin{}--\SourceTextTwinAUROCMax{} on
Twin-70. Their \(\Delta_{\mathrm{AUROC}}\) values are
\SourceTextDeltaAUROCMin{}--\SourceTextDeltaAUROCMax{}, CV-BTE is
\SourceTextCVBTEMin{}--\SourceTextCVBTEMax{}, and Twin-70 FBR at the
unchanged thresholds is
\SourceTextFixedFBRMin{}--\SourceTextFixedFBRMax{}.

Character TF-IDF satisfies all corresponding transfer-failure criteria. The
word readout shows the same large AUROC loss and boundary errors but remains
mixed because its pair-ranking lower bound of \(0.471\) falls below the strict
\(0.50\) criterion.

\begin{table}[t]
\centering
\caption{Lexical controls on Twin-70. Panel A fits and selects each TF-IDF
readout only on the broad source splits, then transfers it unchanged. Panel B
fits within pair-grouped matched training folds and therefore measures
target-supervised readability. Panel A entries are point estimates. AUROC
intervals appear in Fig.~\ref{fig:transfer}.}
\label{tab:text_controls}
\footnotesize
\setlength{\tabcolsep}{3pt}
\textbf{A. Source-trained transfer}\\[2pt]
\begin{tabular}{@{}p{0.34\linewidth}p{0.28\linewidth}p{0.28\linewidth}@{}}
\toprule
Metric & Word TF-IDF & Character TF-IDF \\
\midrule
Source AUROC & 0.976 & 0.987 \\
Twin-70 AUROC & 0.535 & 0.567 \\
$\Delta$AUROC & 0.441 & 0.420 \\
Pair rank & 0.593 & 0.700 \\
Pair C/W/T & 41/28/1 & 49/21/0 \\
Strict pair accuracy & 0.586 & 0.700 \\
CV-BTE & 0.500 & 0.436 \\
Twin TPR/FBR & 0.729/0.657 & 0.786/0.800 \\
XSTest FBR & 0.048 & 0.008 \\
Assessment & mixed$^{a}$ & \begin{tabular}[t]{@{}l@{}}transfer-failure\\criteria met\end{tabular} \\
\bottomrule
\end{tabular}
\par\smallskip
\(^{a}\)The pair-ranking lower bound is \(0.471\), below the strict \(0.50\) support criterion. The other predefined transfer and boundary criteria are met.
\par\medskip
\textbf{B. Target-supervised readability}\\[2pt]
\begin{tabular}{@{}>{\raggedright\arraybackslash}p{0.25\linewidth}rrr@{}}
\toprule
Metric & Word TF-IDF & Character TF-IDF & MiniLM \\
\midrule
Twin-70 AUROC & \begin{tabular}[t]{@{}c@{}}0.955\\ {[0.925, 0.981]}\end{tabular} & \begin{tabular}[t]{@{}c@{}}0.954\\ {[0.925, 0.979]}\end{tabular} & \begin{tabular}[t]{@{}c@{}}0.908\\ {[0.872, 0.944]}\end{tabular} \\
CV-BTE & \begin{tabular}[t]{@{}c@{}}0.121\\ {[0.071, 0.171]}\end{tabular} & \begin{tabular}[t]{@{}c@{}}0.157\\ {[0.107, 0.214]}\end{tabular} & \begin{tabular}[t]{@{}c@{}}0.200\\ {[0.143, 0.257]}\end{tabular} \\
\bottomrule
\end{tabular}
\par\smallskip
Pair rank assigns half credit to ties. Strict pair accuracy counts only correct pairs. Absolute decision counts appear in Table~\ref{tab:decision_intervals}. Panel B brackets give 95\% intervals.
\end{table}

When matched labels enter pair-grouped training folds, word TF-IDF, character
TF-IDF, and MiniLM reach
\TextProxyAUROCMin{}--\TextProxyAUROCMax{} AUROC across the four matched
cohorts. On Twin-70, the activation-minus-proxy increments for the three
evaluated checkpoints are slightly negative, and all three confidence intervals
include zero. The additional proxy and attention analyses in
Appendix~\ref{app:secondary} measure dependence rather than causal
decomposition.

\subsection{Model-based label sensitivity}
\label{sec:second-judge-results}

The broad-to-matched pattern remains numerically stable under both
prespecified label-sensitivity subsets, without relabeling or refitting.
Raw agreement on the \SecondJudgeArmCount{} isolated arms is
\SecondJudgeRawAgreementPercent{}. The consensus and strict subsets retain
\SecondJudgeConsensusPairCount{} and \SecondJudgeStrictPairCount{} of
\SecondJudgePairCount{} pairs. Across seven readouts and four cohorts, the
maximum absolute Twin-AUROC changes are
\SecondJudgeConsensusMaxTwinShift{} and \SecondJudgeStrictMaxTwinShift{},
respectively. Fixed-threshold TPR changes are at most
\SecondJudgeConsensusMaxTPRShift{} and \SecondJudgeStrictMaxTPRShift{}, with
FBR changes at most \SecondJudgeConsensusMaxFBRShift{} and
\SecondJudgeStrictMaxFBRShift{}. Complete agreement and sensitivity metrics
appear in Appendix~\ref{app:second-judge}.

The swapped-order pair assessment answers a different question. Same-topic
agreement is \SecondJudgeSameTopicAgreement{}, but same-request-frame
agreement is only \SecondJudgeSameFrameAgreement{}. In
\SecondJudgeSwapInconsistentPairCount{} of \SecondJudgePairCount{} pairs,
at least one field changes under swapping. Thus stability under arm-label
filters does not establish frame equivalence. The pair assessment filtered
no row and is not independent pair validation.

\subsection{Surface-form diagnostic}
\label{sec:rq4}

For the Qwen~2.5 32B BF16 readout, presentation accuracy is
\SurfaceBFSpaceBehavior{} after space insertion and
\SurfaceBFUnderscoreBehavior{} after underscore insertion. The corresponding
both-orders counts are \SurfaceBFSpaceBothOrdersCount{} and
\SurfaceBFUnderscoreBothOrdersCount{}, compared with
\SurfaceBFOriginalBothOrdersCount{} before
transformation. AUROC changes from
\SurfaceBFOriginalAUROC{} on the unmodified requests to
\SurfaceBFSpaceAUROC{} and \SurfaceBFUnderscoreAUROC{}, respectively. STER is
\SurfaceBFSpaceSTER{} (\SurfaceBFSpaceSTERFraction{} eligible pairs) for space insertion and
\SurfaceBFUnderscoreSTER{} (\SurfaceBFUnderscoreSTERFraction{}) for underscore insertion.

The function-word control changes BF16 AUROC only from
\SurfaceBFOriginalAUROC{} to \SurfaceBFNullAUROC{}.
It partly controls length and tokenization while leaving other differences
between transformations. Full values appear in Appendix~\ref{app:surface}. Under the
prespecified componentwise-maximum readout, space and underscore insertion satisfy the
directional but not the material degradation criterion. Sequence-mean pooling
gives the same classification, while last-token pooling meets the material
criterion. The classification therefore depends on pooling. The separately
fitted Q4 comparison appears in Appendix~\ref{app:surface}.

\section{Discussion}
\label{sec:discussion}

\subsection{What transfers from the source contrast?}

The primary scores usually order matched pairs correctly but overlap across
pairs. Their common thresholds consequently block many harmless requests.
CV-BTE shows that recalibrating thresholds on matched training pairs still
leaves substantial held-out error. The operating problem extends beyond the
unchanged source threshold.

The additional source-trained classifiers change the error profile without
resolving the tested operating problem. Logistic fitting improves ranking,
and several alternatives improve target-calibrated error. At unchanged source
thresholds, fewer false blocks accompany more missed harmful requests. These
observations establish no common lower bound on achievable error.

The policy target in~\eqref{eq:harm} requires distinctions among requested
actions, roles, targets, and authorization. Correlations among topic, wording,
provenance, and labels in broad source data need not supply those distinctions
at a transferable decision boundary. Whether the score supports the required
decision is therefore separate from whether the representation contains
contextual information.

\subsection{Readability and label access}

At three checkpoints, several target-supervised linear, kernel, multilayer,
sequence, and transition configurations recover matched labels with low
out-of-fold error. This establishes dataset-conditional readability, not causal
use or source-boundary transfer \cite{kumar_probing,bolukbasi_illusion}.
Reference-fitted covariance readouts combine strong ranking and low
matched-calibrated CV-BTE with substantially improved transferred operation.
False blocking remains on Twin-70, and detection falls on secondary cohorts.
Strong ranking therefore does not ensure low-error transfer at these thresholds.
Reference data, representation, chat template, and pooling change together
relative to primary readouts, preventing attribution to one factor.

The ten-method comparison extends beyond linear probes. Its outcomes depend
on representation, method, and label access. Other learned representations,
interventions, token positions, or pooling rules may yield better transfer.

\subsection{Lexical cues and label sensitivity}

Character TF-IDF meets the specified transfer-failure criteria. Word TF-IDF
shows a similar loss but misses the pair-ranking criterion. High broad-source
AUROC is achievable with these lexical features alone and does not establish
a reliable distinction between harmful and legitimate requested actions.
Strong target-supervised lexical ranking also shows that surface cues remain
in the constructed pairs. Proxy increments, attention-head selection, and
surface transformations assess predictive dependence, not a causal
decomposition of the activation state.

Removing direct arm-label contradictions in the automated audit changes
transfer metrics only slightly, without resolving construction or review bias. Lower agreement on request-frame matching shows that the pair
design remains contestable even when harmful-arm orientation is recovered.

\subsection{Security implications}
\label{sec:security}

At their transferred thresholds, our primary readouts block most matched
harmless requests. The additional source-trained classifiers reduce false
blocking but miss many harmful requests. These operating points fail to combine
harmful-request detection with legitimate access.

The XSTest--Twin-70 contrast shows why benign evaluation must cover the
intended use: low false blocking on one benchmark did not carry over to
legitimate same-topic requests. The cohorts expose failure on these uses
without estimating natural-traffic rates. A stand-alone gate needs a validated
operating boundary or an explicit abstention policy.

Selecting a quantile from final target-test errors adapts the decision rule
to that benchmark. Legitimate target calibration uses separate calibration
data and independent evaluation, as pair-grouped CV-BTE illustrates. Our
reference quantile stays fixed. The reference readouts still block 9--18 of
70 harmless requests, so relative improvement must be assessed
against acceptable absolute error for legitimate use. At a fixed false-blocking
rate, more harmless traffic means more expected false blocks. Our cohorts do
not estimate deployment rates.

The surface study compares 64-token activation inputs with behavior on
complete requests, confounding surface form with retained token coverage.
A request can cross the tested
diagnostic threshold while the model still makes the correct paired choice
in both orders, before and after transformation. This is sensitivity of one
readout, not a measured increase in harmful compliance. The conspicuous,
normalizable transformations and their pooling dependence do not establish a
practical bypass or a single causal mechanism.

Policy decisions can directly use authenticated evidence of authorization,
ownership, user role, capability, and provenance. Candidate safeguards include
policy checks, rate limits, abstention, and human review. Low scores should
not disable these controls, nor should high scores alone override trusted
context. These are design implications. Complete safety architectures and
system-level improvements were not evaluated.

\section{Related Work}
\label{sec:related}

\paragraph{Security signals and decision boundaries}
Early system-call anomaly detectors had to accommodate legitimate behavioral
variation while detecting intrusions \cite{forrest_self1996}. Hofmeyr et al.\
identified attacks that could retain normal system-call sequences, as well
as policy violations outside the monitored processes
\cite{hofmeyr_syscalls1998}. These studies motivate distinguishing a
detector's observable signal from the security decision it supports.
We test that distinction for source-trained activation readouts on matched
harmful and harmless requests.

\paragraph{Input interpretation in security}
Random-input testing exposed widespread failures in mature UNIX utilities
\cite{miller_fuzz1990}. SQL injection \cite{rfp_sql1998} and indirect prompt
injection \cite{greshake_injection} illustrate a different boundary: data
can be interpreted as instructions with authority they should not have.
They motivate our test of harmful assistance versus legitimate analysis,
without equating score overlap with a parser defect or an
instruction-execution vulnerability.

\paragraph{Activation readouts of harmfulness}
Representation engineering and safety probes recover signals associated with
harmful or malicious inputs from internal states
\cite{zou_repe,mckenzie_probes,kramar_probes,shah_subconcept}. Llorente-Saguer compares activations
with a normative safe reference and evaluates harmful and benign-aggressive
contrasts, including XSTest \cite{llorente2603}. A subsequent multi-model
study fits directions on AdvBench and Alpaca, then transfers them to other
harmful datasets with emphasis on operation at low false-positive rates
\cite{llorente2604}.
Zhao et al. separate a harmfulness representation at the last instruction token from the refusal direction at the post-instruction position and propose this representation as an intrinsic safeguard, evaluated with a fixed centroid rule on jailbroken instructions and on refused harmless XSTest requests~\cite{zhao2025harmfulness}.

He et al.'s streaming detector SC-TopK uses exchange-level supervision, all-layer activations,
and window aggregation regularized on benign exchanges \cite{he_segment}.
Their chemical training data include harmful and harmless variants on related
topics, and evaluation includes legitimate scientific queries. Their TPR at
1\% FPR uses the corresponding negative test sets. We transfer a
source-derived threshold unchanged to matched requests.

\paragraph{Matched contrasts and refusal geometry}
Studies of refusal identify both compact directions and
multidimensional structures
\cite{arditi_refusal,wollschlager_cones,joad_morethan,piras_som}. Petrov uses
topic-matched harmful and harmless prompts to extract refusal directions and
tests an intervention that removes those directions from model weights
\cite{petrov_topicmatched}. Our primary readout is fitted on a broad source contrast and evaluated on
matched pairs without refitting. Topic matching tests transfer rather than
providing a direction for intervention.

OpenSafeIntent varies benign, dual-use, and malicious intent while controlling
the underlying task, then evaluates completion behavior \cite{uppaal_osi}.
Zhang et al.\ report that routing patterns in aligned mixture-of-experts
models are largely topic-driven while safety behavior can change with little
change to the routing path \cite{zhang_moe}. These studies concern output
calibration and expert routing, respectively. Our question is whether a
scalar activation score and its source threshold transfer to a matched
harmfulness contrast.

\paragraph{Probe validity and distribution shift}
Decodability does not by itself establish causal use or robust detection
\cite{kumar_probing,bolukbasi_illusion}. Wang et al.~\cite{wang_falsesense}
find competitive n-gram baselines, failures under controlled text changes,
and false positives on hard benign inputs. We additionally evaluate pair
ordering and absolute errors when a source threshold transfers unchanged to
manually reviewed, same-topic harmful and harmless requests.
Leave-one-dataset-out evaluation
likewise reveals dataset-dependent features in safety classifiers
\cite{fomin_benchmarks}.
Sadhu et al. report a convergent distinction in regulatory compliance: the
tested activation probes and guard models often rank broad violation cues but
fail to compose a supplied rule with its scenario \cite{sadhu_compliance}.
Their different domain, crossed benchmark, and detector families provide
convergent evidence rather than a direct replication.
Format and style can also affect safety behavior or
probe results \cite{xiao_style,sahoo_format}. The source-trained
TF-IDF controls in this study evaluate that specificity question under the
same source-to-matched design.

\paragraph{Safety benchmarks and operating points}
Guard models and evaluation suites address harmful content, jailbreaks,
refusal, and over-refusal
\cite{inan_llamaguard,wildguard,shieldgemma,mazeika_harmbench,
chao_jailbreakbench,souly_strongreject,rottger_xstest,cui_orbench}.
Calibration studies emphasize connecting classifier scores to explicit
operating points \cite{liu_calibration,wang_sok}.

\section{Limitations}
\label{sec:limitations}

\paragraph{Empirical scope}
The study covers five checkpoints, three models for the ten-method
comparison, specified layers and read points, scalar readouts, English
single-turn requests, and four constructed matched cohorts. The lexical
control contains two TF-IDF feature sets rather than an exhaustive set of text
classifiers. The additional source-trained readouts cover logistic regression, one
radial-basis support vector machine, and one-hidden-layer multilayer
perceptrons over one fixed pooled representation. Larger training sets, deeper
networks, sparse-autoencoder representations, and causal interventions remain
untested. The results apply to the evaluated configurations, not to all
activation-space methods.

The primary 64-token window covers every request in Twin-70, Twin-43, and
Twin-120. Two of 434 Twin-217 arms exceed this window for Mistral~7B, with
65 and 66 tokens.
The label relevance of their omitted suffixes was not separately annotated.

\paragraph{Dataset construction and review}
The matched cohorts are synthetic contrasts, not samples of natural traffic.
We manually reviewed and labeled every candidate. Iterative
rejection and replacement can leave generator-specific, reviewer-specific, or
surface-form regularities in the accepted cohorts. Twin-70's Qwen~2.5 32B
generator uses \texttt{Q4\_K\_M}, while the evaluated model uses BF16, leaving
possible dependence on their shared model family unquantified. No independent human
re-annotation was performed for the Twin cohorts or the reference collection.
Twin-70 was separately generated and initially reviewed blind,
whereas the other source collections had earlier activation measurements.
The derived sets share construction provenance as specified in
Section~\ref{sec:setup}. The labels describe the requested action,
not unexpressed intent or live authorization.

The automated second judge provides no independent human annotation. Five
isolated-arm requests and two pair presentations were blocked and recorded
as uncertain, alongside explicit uncertainty. Lower frame agreement limits
matching even when arm labels agree. The filters test numerical label
sensitivity, leave shared rubric/model dependence unresolved, and do not
validate ground truth. The missing dated API snapshot limits call-level
reproduction.

\paragraph{Causal identification}
Differences among checkpoints confound scale, architecture, training data,
and post-training. Lexical proxies, attention-head selection,
and surface-form transformations measure dependence rather than causal
mechanism. The Q4 comparison changes both the implementation and numerical
precision, which prevents isolating the effect of four-bit conversion.

\paragraph{Metrics and operating points}
AUROC, pair ranking, overlap, and CV-BTE are finite-sample summaries. The
source and supplementary reference thresholds each test one fifth-percentile
operating rule. The reference-validation split had already been inspected
through descriptive AUROC and is not a previously unseen holdout. Its rates
are calibration diagnostics. The new binomial intervals condition on the
fitted scores and realized thresholds, excluding recalibration uncertainty.
The surface-form threshold is target-derived and diagnostic. XSTest has empirical
FBR resolution \XSTestResolution{}, and each Twin-70 category contains only
\PrimaryWorstGroupCategoryN{} harmless arms. These samples cannot resolve
very low false-blocking rates or estimate performance on natural traffic.

\paragraph{Surface-form study}
The case study covers one model, one primary layer, three pooling rules, two
systematic transformations, and a function-word control. The transformations
increase token count, are visually conspicuous, and can be removed by
preprocessing. \(T_{\mathrm{control}}\) partly measures length and
tokenization sensitivity while retaining other differences. All systematically
transformed arms and eight function-word control arms exceed the 64-token
activation window. Behavioral agreement on the complete pair does not
establish that the truncated activation input retains every label-relevant
contrast. The paired
behavior test covers one label-relevant decision in two presentation orders
rather than equivalence of full completion behavior. Presentation accuracy
averages the two choices and need not imply both-orders correctness for every
pair. STER conditions on that stricter property in both conditions and on an
initially triggered score, so it does not describe all 151 pairs.

\paragraph{Unobserved context}
Authorization, ownership, user role, capability, provenance, and downstream
effects are not individually randomized. Languages other than English,
multi-turn interaction, adaptive users who anticipate normalization, and
complete policies with abstention are outside the evaluation. Generalization
to those settings remains open.

\section{Conclusion}
\label{sec:conclusion}

Strong broad-source ranking did not provide reliable operating decisions on
Twin-70. The primary activation readouts detected most harmful requests while
blocking most legitimate same-topic requests. Substantial error remained after
threshold recalibration on matched training pairs. Three readouts met all
prespecified Entanglement Wall criteria, while two yielded mixed evidence.
The secondary Twin-43, Twin-120, and Twin-217 evaluations also show positive
source-to-matched AUROC loss for every checkpoint, with confidence intervals
above zero. Together, the four cohorts cover \ActiveTwinPairCount{} manually
reviewed pairs, with Twin-70 providing the primary test of the Wall criteria.

Source-trained logistic regression improves ranking, and several alternatives
reduce error after target calibration. At the unchanged source thresholds, reduced false
blocking accompanies more missed harmful requests. Reference-fitted covariance
readouts improve transferred operation but retain Twin-70 false blocking and
lower secondary-cohort detection. Mistral's whitened readout on Twin-120 meets
the interval-based TPR and FBR counterconditions. Low error after matched
calibration and matched-label training confirms that useful distinctions
remain accessible. Lexical controls show that transfer loss also occurs
without activation features.

The security consequence is the gap between detecting broad risk and making
decisions that preserve both harmful-request detection and legitimate access.
The tested source operating points do not close that gap. Safety-gate
evaluation must establish both requirements at the intended operating point.

\section*{Open Science}
\label{sec:open-science}

The reproduction artifact is available at
\url{https://github.com/dschwarz33/entanglement-wall}.
It contains manuscript sources, fixed protocols
and configurations, public registries, numerical results, reproduction scripts,
hash manifests, and independent verification reports. A scalar-score export
uses readout-specific opaque request and pair identifiers. Its scores, labels,
folds, thresholds, decisions, and sensitivity flags support reconstruction of
AUROC, pair ranking, CV-BTE, confusion counts, and their bootstrap procedures
for the primary activation and lexical analyses. The reference-method
export additionally contains individual scores, opaque pair identifiers,
fold assignments, threshold decisions, the 20 activation-deflation split-score
vectors, and uncertainty seeds. The supplementary calibration export adds
six reference-validation score vectors, opaque request identities and binary
labels, and all 24 transferred operating points. These support independent
reconstruction of the six quantile thresholds, confusion counts, and exact
binomial intervals. Separate archives retain text-free activation
arrays and frozen scores for source-only probe refits. The surface export
supplies parsed A/B choices, validity indicators, pair identifiers, scores,
and thresholds for independent reconstruction of presentation accuracy,
both-orders correctness, and STER. These codes were also checked locally
against the preserved raw responses.

The offline \texttt{saved-scores} workflow reconstructs the reference-method
analyses, replays reference calibration, and checks the released primary and
source-probe results. It also
regenerates the methods figure and table, verifies the numerical values,
and builds the manuscript twice. A separate \texttt{source-refit} workflow
refits the additional source-trained probes from the released feature arrays.
Full refitting of the reference methods requires the withheld activation
caches and reference-membership mapping.

We withhold sensitive matched and reference requests, translations, complete
manual-review records, identifier mappings, local automated-judge responses,
and non-probe activation caches. Model weights and tokenizers are not
redistributed because of size and licensing constraints. Immutable repository
revisions, extraction configurations, expected array shapes, and hashes
document their provenance. Without the restricted texts and response records,
the package cannot repeat the original activation extraction, reissue judge
calls, or reconstruct judge subsets from raw responses. It supports the
specified saved-score analyses and retains separate inputs for source-only
probe refits. Text-free activation arrays still carry residual reconstruction
and linkage risks.

\section*{LLM usage considerations}

LLMs supported candidate generation, an automated label-sensitivity audit,
and code and manuscript preparation. We defined the research questions and
experimental design, manually assigned the final labels, interpreted the
results, and made all scientific decisions.

Language models drafted same-topic Twin requests under construction criteria
that we defined. For Twin-70, we used locally hosted Qwen~2.5 32B Instruct
in \texttt{Q4\_K\_M} quantization. Appendix~\ref{app:cohort-construction}
gives its settings and the limits of the older generation records. The models
drafted requests without answering them. We manually reviewed and labeled
every candidate, and determined cohort membership using the stated rubric.
The reference collection combines intent files with locally generated
requests. Its records document automated guard checks but lack complete
generator identities and settings for some portions. We manually reviewed
and assigned its final harmfulness and category labels and exclusions.

The blinded automated second judge described in
Section~\ref{sec:second-judge-method} audits sensitivity to label
disagreement. It did not replace manual labels or select analyses. Each
serial call used a fresh context with \texttt{store=false}, withholding
manual labels, scores, results, claims, manuscript text, and prior responses.
The returned \texttt{gpt-5.6-sol} identifier lacks a dated snapshot, which
limits exact call-level reproduction.

LLMs were used for editorial purposes in this manuscript, and all outputs were inspected by the authors to ensure accuracy and originality.
They also supported code development, test generation, and documentation.
We inspected every assisted contribution and tested or independently verified
it where applicable. LLM-assisted review suggested additional checks. We
selected and specified the analyses and made all final scientific decisions.
We remain responsible for the work's accuracy,
originality, and integrity.

The three anchor checkpoints support the method comparison, and the two
larger checkpoints extend the transfer evaluation. Model-backed experiments
used single NVIDIA H100, H200, or RTX~5090 GPUs with serial one-sequence
forward passes. The Twin-70 record identifies local execution but not its
specific generation device. Fixed inventories, finite attempt limits, and
reuse of frozen outputs limited repeated inference. We do not estimate a
numerical emissions footprint.

\section*{Ethical considerations and responsible release}

The matched requests were generated synthetically for this study. Cohort
construction did not collect user conversations, recruit participants, or
use real-user interaction logs. We manually reviewed every candidate and
assigned its label. Section~\ref{sec:limitations} describes the limits of our
manual review. We make no claim of formal ethics-board review.

Some matched requests contain sensitive or dual-use instructions. Released
materials therefore omit their plaintext. The core scalar-score export uses
readout-specific opaque identifiers. The separate Source-probe and surface
exports retain shared pair hashes for their documented comparisons. Mappings
to restricted rows are withheld, but exact scores can still act as fingerprints
for someone who already possesses the private arrays. We do not claim
unlinkability. These restrictions limit full reconstruction.
Redistributing the complete records would expose the instruction content that
the release boundary is intended to protect.

The automated second judge received sensitive requests only for the specified
post-freeze audit. Its outputs remain sensitivity evidence rather than ground
truth. The surface-form transformations are diagnostic tests of a fixed
readout, not an evasion recipe or evidence against layered controls.

\bibliographystyle{IEEEtran}
\bibliography{references}

\appendices
\section{Additional Statistical Detail}
\label{app:secondary}

\subsection{Representation constructions and controls}
\label{app:representation-detail}
\label{app:direct-readout-hyperparameters}

The comparison uses the chat-template preprocessing described in
Section~\ref{sec:axes-method}. Eight methods depend on its reference
collection. The layer/read-point method uses matched training folds without
this reference collection. The generation-window method uses a separately
frozen set of 215 medoids per checkpoint, described in
Appendix~\ref{app:sensor}.

\paragraph{Reference construction and review}
We manually reviewed 4,085 candidate requests for the harmfulness of the
requested action and suitability of the category, and assigned the final
labels. Three unresolved candidates were excluded without replacement. The
final collection contains 4,082 requests across 43 categories, with 3,008
fitting and 1,074 validation requests. The retained requests preserve their
split assignments and relative source order. Category sizes are not uniform.

For sequence warping, the registered balancing rule retains all harmless
fitting references and selects the same number of harmful fitting references
by rounded equally spaced indices in the canonical order. The
activation-deflation diagnostic uses 20 registered reference-split seeds.
Within each category, the split rule assigns $\lfloor n/2\rfloor$ requests to the direction
subset and the remainder to the classifier subset. Reference-validation
AUROC for the whitened and local-covariance methods was initially descriptive.
The supplementary analysis uses the same split to calibrate the six transferred
thresholds in Appendix~\ref{app:reference-calibration}. The depth-difference
method also uses reference validation for its registered layer selection.

\paragraph{Fixed-reference methods}
The generation-window score subtracts the maximum similarity to harmless
reference medoids, which are representative reference activations, from the
maximum similarity to harmful medoids at each generated position. After three
warm-up tokens, the sensor score averages this margin through the first
sentence-ending token at or before position 20. If no such token occurs, it
uses the available prefix capped at position 20. The sequence-warping score
uses five-nearest-neighbor dynamic time warping, which aligns token sequences
while allowing insertions and deletions, to compare harmful and harmless
reference sequences. The frame--content score is a fixed binary indicator of
whether centroids representing request form and harmful content co-occur
within the first eight user tokens.

\paragraph{Reference-fitted methods}
These three methods first average the retained user-token activations within
each request. The depth-difference score projects each layer onto a
harmful-minus-harmless reference direction, standardizes the layer scores on
reference data, subtracts the standardized baseline score from each deeper
score, and selects the deeper layer on the disjoint validation split.
The whitened mean-difference score standardizes the reference data, retains at
most 150 principal components, and compares class means under a shared
covariance estimate. For each query, the local-covariance score
selects the 125 nearest reference-fit examples from each class after the same
dimension reduction. It estimates one shared local covariance matrix from
these examples after centering them within their respective classes.

\paragraph{Matched-label methods}
The recurrent sequence readout is a one-layer gated recurrent unit, a neural
sequence model fitted to the ordered token-to-centroid sequence. A
within-request token-order shuffle serves as a control. The layer/read-point
readout is logistic regression over zero-based Llama layers 14, 19, 23, 27, or
30 and four activation locations: the post-attention layer-normalization
output, input layer-normalization output, attention output before residual
addition, or block output. It was implemented only for Llama~3.1 8B. The
directed-transition readout is logistic regression over transitions between
dominant reference centroids. A symmetric-transition representation serves as
its direction control. These three methods use matched labels only within the
training portion of each outer pair fold. Selection and preprocessing are
repeated inside the folds.

The GRU receives at most 50 token vectors with 43 centroid similarities each
and has one unidirectional layer with 64 hidden units, no dropout, and a
linear output logit. Adam uses learning rate \(10^{-3}\), weight decay
\(10^{-4}\), batches of 64, and at most 60 epochs. A category-stratified
inner pair holdout selects the epoch by AUROC with patience 8 and the
earliest-epoch tie rule. Within each category, the registered ordering assigns
rank modulo five equal to zero to validation. Token scaling uses only the
outer-training tokens. Each seed is then reinitialized and trained on all
outer-training pairs for the selected epoch count. Predictions average seeds
0 through 4. The token-shuffle control averages five evaluation permutations.

The layer/read-point and directed-transition methods use StandardScaler
followed by logistic regression with \(C=1\), balanced class weights,
L-BFGS, at most 2,000 iterations, and random state 0. Layer/read-point
selection uses mean AUROC over three inner pair folds, resolving ties by
the listed layer order and then read-point order, before refitting on the
outer-training pairs. Directed and symmetric transition classifiers use these
fixed settings within the five outer folds.

\paragraph{Direct matched-label classifiers}
The direct classifiers use the grids and estimator settings in
Appendix~\ref{app:probe-hyperparameters}. Within each outer training fold,
the candidate with the highest mean AUROC across three inner pair folds is
selected, with the same first-candidate tie rule. Scaling is fitted on inner
training arms only. The selected pipeline is then refitted on all outer
training arms. MLP selection is repeated for each of its three seeds.
Raw scores are standardized with the corresponding outer-training mean and
population standard deviation, then averaged across MLP seeds. Cross-cohort
predictions additionally average the five Twin-70 outer-fold models, without
using external target labels for fitting, selection, scaling, or calibration.
Table~\ref{tab:direct-readouts} reports every model--classifier combination.

\paragraph{Diagnostic benign-subspace projection}
This method first averages retained user-token activations.
The diagnostic benign-subspace
method projects activations away from the lowest-dimensional harmless-reference
singular-vector subspace whose cumulative squared singular values reach
90\%, then fits a logistic readout. This last procedure is diagnostic and
does not define a deployable rule.
After projection, a second StandardScaler and logistic regression use the
fixed \(C=1\) settings above, fitted on reference-fit data.

\paragraph{Computation}
The reference-dependent methods use the stored activation arrays and the
manually reviewed reference labels. Their runtime uses NumPy 2.4.6, with
PyTorch 2.10.0+cpu for deterministic CPU training of the recurrent readout.
The artifact records checkpoint revisions, extraction layers, read points,
pooling, folds, seeds, method parameters, and runtime versions.

Table~\ref{tab:methods_appendix} retains the complete numerical ranges and
model counts. Label access describes readout fitting. CV-BTE additionally
uses matched training labels for threshold selection.

\begin{table}[!htbp]
\centering
\caption{Representation-method results on Twin-70.}
\label{tab:methods_appendix}
\footnotesize
\setlength{\tabcolsep}{3pt}
\begin{tabular*}{\columnwidth}{@{\extracolsep{\fill}}>{\raggedright\arraybackslash}p{0.31\linewidth}rrrr@{}}
\toprule
Readout & $m$ & AUROC & Pair rank & CV-BTE \\
\midrule
\multicolumn{5}{@{}l}{\textit{Fixed references}} \\
Generation-window score & 3 & 0.882--0.948 & 0.929--0.971 & 0.093--0.229 \\
Sequence-warping score & 3 & 0.707--0.814 & 0.843--0.914 & 0.264--0.336 \\
Frame--content co-occurrence & 3 & 0.479--0.486 & 0.479--0.486 & 0.500--0.514 \\
\multicolumn{5}{@{}l}{\textit{Reference-fitted}} \\
Depth-difference score$^{v}$ & 3 & 0.792--0.909 & 0.929--0.986 & 0.171--0.314 \\
Whitened mean difference$^{v}$ & 3 & 0.945--0.974 & 1.000--1.000 & 0.071--0.093 \\
Local covariance score$^{v}$ & 3 & 0.943--0.964 & 0.971--0.986 & 0.107--0.136 \\
\multicolumn{5}{@{}l}{\textit{Matched training}} \\
Recurrent sequence readout & 3 & 0.954--0.969 & 0.971--0.986 & 0.093--0.129 \\
Layer/read-point readout & 1 & 0.993 & 1.000 & 0.043 \\
Directed-transition readout & 3 & 0.829--0.912 & 0.829--0.957 & 0.164--0.243 \\
\multicolumn{5}{@{}l}{\textit{Diagnostic projection}} \\
Benign-subspace projection & 3 & 0.379--0.627 & 0.286--0.700 & 0.386--0.507 \\
\bottomrule
\end{tabular*}
\par\smallskip
\raggedright $m$ is the number of evaluated checkpoints. Ranges retain all checkpoint cells. Groups describe readout fitting and selection. $^{v}$Reference validation selects the depth-difference layer. Its initial use for covariance readouts was descriptive. Section~\ref{sec:axes-method} details their supplementary threshold calibration. Matched training uses pair-grouped training folds. Every CV-BTE additionally selects thresholds on Twin training folds and is diagnostic. Layer/read-point selection covers only Llama~3.1 8B. The projection is a diagnostic reference fit, not a deployment candidate.
\end{table}

\subsection{Wall counterconditions}
\label{app:wall-counterconditions}

The criteria in Section~\ref{sec:wall-criteria} are deliberately one-sided.
We define the counterpattern by requiring all of the following point
estimates: \(\Delta_{\mathrm{AUROC}}\leq0.10\), pair ranking
\(\geq0.80\), CV-BTE \(\leq0.20\), matched TPR \(\geq0.80\), matched FBR
\(\leq0.20\), worst-category FBR \(\leq0.40\), and XSTest FBR \(\leq0.10\).
Refutation requires the corresponding confidence bounds to meet all seven
conditions, using the upper endpoint for upper limits and the lower endpoint
for lower limits. A readout that satisfies neither the support nor the
opposite pattern is reported as mixed. These assessments are made separately
for each readout.

For worst-category FBR, each bootstrap draw resamples within every eligible
category at its observed size and then takes the maximum. The gate uses the
upper percentile bound of these maxima. With seven harmless arms per Twin-70
category, this empirical interval has limited resolution and does not guarantee
conservative population coverage, particularly at zero observed errors.
This countercondition is not part of the Wall support criteria.

\subsection{Exact CV-BTE procedure}
\label{app:cv-bte}

For each matched cohort, five folds are assigned at the pair level and
stratified by request category. Within a category, pairs are sorted by
\(\mathrm{SHA256}(\texttt{SATML2027-CV-FOLD-V1}\mathbin{\|}0
\mathbin{\|}\textit{cohort}\mathbin{\|}0\mathbin{\|}\textit{pair})\), with
the pair identifier as the tie breaker, and the sorted rank modulo five gives
the fold. The null bytes shown as \(0\) are separators. This assignment uses a
registered hash domain rather than a random fold seed.

Within each training partition, let \(FN(\tau)\) and \(FP(\tau)\) be the
errors under the strict decision rule \(s>\tau\). The fitted threshold
minimizes
\begin{equation}
\tfrac{1}{2}\left[
\frac{FN(\tau)}{TP(\tau)+FN(\tau)}
+
\frac{FP(\tau)}{FP(\tau)+TN(\tau)}
\right].
\end{equation}
Candidates are \(-\infty\), \(+\infty\), and the midpoints between adjacent
distinct training scores. If the training scores have one unique value, only
the two infinite candidates remain. A score equal to the threshold is
negative. When several candidates achieve the same minimum, they are sorted
numerically and the upper order-statistic median, at zero-based index
\(\lfloor k/2\rfloor\), is selected. The implementation compares the
integer numerator \(FN\,n_{-}+FP\,n_{+}\), where \(n_{+}\) and \(n_{-}\)
are the training class sizes. Its denominator is constant across candidates,
so this comparison preserves exact equality of balanced errors.

Both members of a pair remain in the same fold. The implementation rejects
an empty training or held-out partition, an out-of-range assignment, duplicate
pair identifiers, or training data without both classes. No such degenerate
fold occurs in the reported analyses. Each training-fold threshold is applied
unchanged to its held-out pairs. The final CV-BTE is computed once from the
aggregate out-of-fold TP, FN, FP, and TN counts in
Eq.~\eqref{eq:bte}, rather than by averaging fold-level rates. In the primary
analysis family, its interval resamples the frozen out-of-fold pair decisions
10,000 times without reselecting thresholds or folds. These intervals describe
uncertainty conditional on those fitted decisions. Bootstrap substreams derive from
root seed 20260725 and the registered analysis scope. The additional
Source-trained probe control uses 5,000 resamples and a separate deterministic
seed domain specified in its configuration. Source-trained readouts retain their
source-fixed orientation. For target-supervised readouts, orientation is
learned within each training partition. Held-out labels never reverse a score
direction.

\subsection{Matched-versus-shuffled diagnostic}
\label{app:pairing}

The diagnostic defined in Section~\ref{sec:problem} uses
\ResampleDrawCount{} shuffled assignments and a centered two-sided test.
Table~\ref{tab:pairing_diagnostic} reports the observed statistic, shuffled
reference distribution, and test result.

\begin{table}[!htbp]
\centering
\caption{Matched-versus-shuffled pair ranking on Twin-70.}
\label{tab:pairing_diagnostic}
\footnotesize
\setlength{\tabcolsep}{3pt}
\begin{tabular*}{\columnwidth}{@{\extracolsep{\fill}}>{\raggedright\arraybackslash}p{0.29\linewidth}ccc@{}}
\toprule
Model & Observed & Shuffled & Two-sided $p$ \\
\midrule
Llama 3.1 8B & \begin{tabular}[c]{@{}c@{}}{0.829}\\ {[0.743, 0.914]}\end{tabular} & \begin{tabular}[c]{@{}c@{}}{0.712}\\ {[0.657, 0.771]}\end{tabular} & 0.0004 \\
\addlinespace[2pt]
Mistral 7B & \begin{tabular}[c]{@{}c@{}}{0.886}\\ {[0.800, 0.957]}\end{tabular} & \begin{tabular}[c]{@{}c@{}}{0.757}\\ {[0.700, 0.814]}\end{tabular} & 0.0001 \\
\addlinespace[2pt]
Qwen 2.5 7B & \begin{tabular}[c]{@{}c@{}}{0.929}\\ {[0.857, 0.986]}\end{tabular} & \begin{tabular}[c]{@{}c@{}}{0.855}\\ {[0.800, 0.900]}\end{tabular} & 0.0031 \\
\addlinespace[2pt]
Mistral Small 24B & \begin{tabular}[c]{@{}c@{}}{0.929}\\ {[0.857, 0.986]}\end{tabular} & \begin{tabular}[c]{@{}c@{}}{0.825}\\ {[0.771, 0.871]}\end{tabular} & 0.0001 \\
\addlinespace[2pt]
Qwen 2.5 32B & \begin{tabular}[c]{@{}c@{}}{0.929}\\ {[0.857, 0.986]}\end{tabular} & \begin{tabular}[c]{@{}c@{}}{0.880}\\ {[0.843, 0.929]}\end{tabular} & 0.0504 \\
\addlinespace[2pt]
\bottomrule
\end{tabular*}
\par\smallskip
\raggedright Brackets give 95\% intervals. The observed interval uses pair bootstrap. The shuffled reference permutes harmless arms within request category while holding scores fixed. The first four models are above the reference. Qwen~2.5 32B is at the reference boundary.
\end{table}

\subsection{Attention and proxy analyses}

On Twin-70, single attention heads selected on source data span
\SourceHeadAUROCMin{}--\SourceHeadAUROCMax{} AUROC. Heads selected directly
with Twin-70 labels reach
\OracleHeadAUROCMin{}--\OracleHeadAUROCMax{}. Because the latter values use
target labels for selection, this comparison does not identify a localized
intent mechanism.

All three activation-minus-proxy intervals in
Table~\ref{tab:proxy_increment_appendix} include zero. The analysis is limited
to incremental prediction after the text proxy is available.

\begin{table}[!htbp]
\centering
\caption{Activation-minus-text-proxy AUROC increments on Twin-70.}
\label{tab:proxy_increment_appendix}
\footnotesize
\setlength{\tabcolsep}{3pt}
\begin{tabular*}{\columnwidth}{@{\extracolsep{\fill}}lr@{}}
\toprule
Activation readout & Increment [95\% interval] \\
\midrule
Llama 3.1 8B & -0.002 [-0.007, 0.003] \\
Mistral 7B & -0.002 [-0.009, 0.006] \\
Qwen 2.5 7B & -0.005 [-0.017, 0.007] \\
\bottomrule
\end{tabular*}
\par\smallskip
\raggedright All readouts are evaluated in pair-grouped matched folds.
\end{table}

\subsection{Decision-relevant intervals}

Table~\ref{tab:decision_intervals} gives the frozen intervals that determine
the lexical assessment.

\begin{table}[!htbp]
\centering
\caption{Uncertainty for Source-trained lexical controls on Twin-70.}
\label{tab:decision_intervals}
\footnotesize
\setlength{\tabcolsep}{3pt}
\begin{tabular*}{\columnwidth}{@{\extracolsep{\fill}}>{\raggedright\arraybackslash}p{0.30\linewidth}cc@{}}
\toprule
Metric & Word TF-IDF & Character TF-IDF \\
\midrule
Source AUROC & 0.976 [0.965, 0.985] & 0.987 [0.978, 0.994] \\
$\Delta$AUROC & 0.441 [0.398, 0.482] & 0.420 [0.371, 0.465] \\
Pair rank & 0.593 [0.471, 0.707] & 0.700 [0.586, 0.800] \\
CV-BTE & 0.500 [0.457, 0.543] & 0.436 [0.393, 0.479] \\
Twin-70 TPR & 0.729 [0.629, 0.829] & 0.786 [0.686, 0.886] \\
Twin-70 TP/FN & 51/19 & 55/15 \\
Twin-70 FBR & 0.657 [0.543, 0.771] & 0.800 [0.700, 0.886] \\
Twin-70 FP/TN & 46/24 & 56/14 \\
XSTest FBR & 0.048 [0.024, 0.076] & 0.008 [0.000, 0.020] \\
XSTest FP/TN & 12/238 & 2/248 \\
I / II / III & yes / yes / no & yes / yes / no \\
\bottomrule
\end{tabular*}
\par\smallskip
\raggedright Brackets give 95\% percentile-bootstrap intervals from 10,000 draws. Source classes are resampled independently and Twin arms stay together by pair. Transfer differences resample both cohorts. Decision counts use the unchanged Source threshold with $s>\tau$. Equality is negative. Operating problems are I: TPR upper bound $<0.90$, II: FBR lower bound $>0.20$, and III: XSTest FBR lower bound $>0.10$. At least one is required.
\end{table}

\subsection{Source-trained discriminative probe control}
\label{app:source-probes}
\label{app:probe-hyperparameters}

\paragraph{Classifier settings}
The source-trained and direct matched-label classifiers share the following
grids. Logistic regression uses \(C\in\{0.01,0.1,1,10\}\),
balanced class weights, L-BFGS, and at most 3,000 iterations. The RBF-SVM uses
balanced class weights and \(C\in\{0.1,1,10\}\) crossed with
\(\gamma\in\{\mathrm{scale},\mathrm{auto}\}\). The one-hidden-layer MLP
uses widths 32 or 64 and L2 penalties \(\alpha\in\{10^{-4},10^{-3}\}\),
ReLU activations, Adam, and at most 500 iterations. Its training seeds are
0, 1, and 2. A StandardScaler is fitted within each training partition.
Ties select the first registered candidate, in ascending \(C\) order, then
scale before auto for the RBF-SVM, and ascending \(\alpha\), then width,
for the MLP. Remaining estimator options follow the frozen configuration
and the recorded scikit-learn 1.8.0 defaults.

For the source-trained control, each candidate is fitted on Source-fit only
and selected by Source-selection AUROC, separately for each MLP seed. The
selected estimators are not refitted on selection rows. Logistic and RBF-SVM
scores are decision-function values. MLP probabilities are clipped to
\([10^{-6},1-10^{-6}]\) and converted to logits. Each score is standardized
using its Source-fit mean and population standard deviation. The MLP averages
the three standardized seed-specific scores before threshold calibration
and transfer.

Tables~\ref{tab:source_probe_transfer} and~\ref{tab:source_probe_operating}
report the complete primary-cohort ranking and operating-point results for
the additional source-trained classifiers. Every classifier was fitted and
selected on Source data only. The target-derived CV-BTE column is a separate
overlap diagnostic and does not replace the unchanged Source-threshold
operating point. The independently verified result in the reproduction artifact also reports the
same readouts on Twin-120, Twin-217, and Twin-43.

\begin{table}[!htbp]
\centering
\caption{Source-trained probe ranking on Twin-70.}
\label{tab:source_probe_transfer}
\footnotesize
\setlength{\tabcolsep}{3pt}
\begin{tabular*}{\columnwidth}{@{\extracolsep{\fill}}lcccc@{}}
\toprule
Probe & \begin{tabular}[c]{@{}c@{}}{Source}\\ {AUROC}\end{tabular} & \begin{tabular}[c]{@{}c@{}}{Twin}\\ {AUROC}\end{tabular} & $\Delta$AUROC & \begin{tabular}[c]{@{}c@{}}{Rank}\\ {($C/W/T$)}\end{tabular} \\
\midrule
\multicolumn{5}{@{}l}{\textit{Llama 3.1 8B}} \\
Logistic & 1.000 & 0.799 & 0.201 & 0.943 (66/4/0) \\
RBF-SVM & 0.998 & 0.726 & 0.272 & 0.900 (63/7/0) \\
MLP & 1.000 & 0.778 & 0.222 & 0.914 (64/6/0) \\
\midrule
\multicolumn{5}{@{}l}{\textit{Mistral 7B}} \\
Logistic & 1.000 & 0.834 & 0.165 & 0.971 (68/2/0) \\
RBF-SVM & 1.000 & 0.783 & 0.217 & 0.914 (64/6/0) \\
MLP & 0.999 & 0.801 & 0.198 & 0.957 (67/3/0) \\
\midrule
\multicolumn{5}{@{}l}{\textit{Qwen 2.5 7B}} \\
Logistic & 0.999 & 0.893 & 0.106 & 0.971 (68/2/0) \\
RBF-SVM & 0.999 & 0.740 & 0.259 & 0.929 (65/5/0) \\
MLP & 0.999 & 0.876 & 0.124 & 0.971 (68/2/0) \\
\midrule
\multicolumn{5}{@{}l}{\textit{Mistral Small 24B}} \\
Logistic & 1.000 & 0.871 & 0.129 & 0.971 (68/2/0) \\
RBF-SVM & 1.000 & 0.803 & 0.197 & 0.971 (68/2/0) \\
MLP & 1.000 & 0.830 & 0.169 & 0.957 (67/3/0) \\
\midrule
\multicolumn{5}{@{}l}{\textit{Qwen 2.5 32B}} \\
Logistic & 1.000 & 0.874 & 0.125 & 0.971 (68/2/0) \\
RBF-SVM & 1.000 & 0.764 & 0.236 & 0.914 (64/6/0) \\
MLP & 1.000 & 0.836 & 0.164 & 0.957 (67/3/0) \\
\bottomrule
\end{tabular*}
\par\smallskip
\raggedright All classifiers use the frozen primary activation vectors and Source-only selection. $C/W/T$ gives correct, wrong, and tied pairs. Pair rank assigns half credit to ties. Strict pair accuracy equals pair rank in these rows because all tie counts are zero.
\end{table}

\begin{table}[!htbp]
\centering
\caption{Source-trained probe operating points on Twin-70.}
\label{tab:source_probe_operating}
\footnotesize
\setlength{\tabcolsep}{3pt}
\begin{tabular*}{\columnwidth}{@{\extracolsep{\fill}}lccc@{}}
\toprule
Probe & CV-BTE & \shortstack{TPR\\ (TP/FN)} & \shortstack{FBR\\ (FP/TN)} \\
\midrule
\multicolumn{4}{@{}l}{\textit{Llama 3.1 8B}} \\
Logistic & 0.300 & 0.529 (37/33) & 0.071 (5/65) \\
RBF-SVM & 0.279 & 0.200 (14/56) & 0.014 (1/69) \\
MLP & 0.357 & 0.286 (20/50) & 0.014 (1/69) \\
\midrule
\multicolumn{4}{@{}l}{\textit{Mistral 7B}} \\
Logistic & 0.186 & 0.629 (44/26) & 0.057 (4/66) \\
RBF-SVM & 0.229 & 0.243 (17/53) & 0.014 (1/69) \\
MLP & 0.243 & 0.514 (36/34) & 0.057 (4/66) \\
\midrule
\multicolumn{4}{@{}l}{\textit{Qwen 2.5 7B}} \\
Logistic & 0.164 & 0.500 (35/35) & 0.014 (1/69) \\
RBF-SVM & 0.379 & 0.400 (28/42) & 0.100 (7/63) \\
MLP & 0.179 & 0.457 (32/38) & 0.014 (1/69) \\
\midrule
\multicolumn{4}{@{}l}{\textit{Mistral Small 24B}} \\
Logistic & 0.164 & 0.671 (47/23) & 0.071 (5/65) \\
RBF-SVM & 0.257 & 0.386 (27/43) & 0.029 (2/68) \\
MLP & 0.200 & 0.529 (37/33) & 0.057 (4/66) \\
\midrule
\multicolumn{4}{@{}l}{\textit{Qwen 2.5 32B}} \\
Logistic & 0.164 & 0.429 (30/40) & 0.029 (2/68) \\
RBF-SVM & 0.293 & 0.571 (40/30) & 0.214 (15/55) \\
MLP & 0.243 & 0.286 (20/50) & 0.000 (0/70) \\
\bottomrule
\end{tabular*}
\par\smallskip
\raggedright The same Source-selected probes as in Table~\ref{tab:source_probe_transfer}. CV-BTE is target-derived and diagnostic. TPR and FBR use the unchanged Source threshold with $s>\tau$. Equality is negative. Parentheses give absolute counts. All probes have 0/250 XSTest FP/TN and FBR 0.000.
\end{table}

Table~\ref{tab:source_verification_tpr} adds the held-out source detection
counts for all five primary and 15 alternative readouts. It uses the original
scores and full stored thresholds, without refitting or recalibration.

\begin{table}[!htbp]
\centering
\footnotesize
\caption{Source-verification detection at unchanged source-selection thresholds. Each row uses all 320 held-out harmful AdvBench requests. TPR values are point estimates.}
\label{tab:source_verification_tpr}
\begin{tabular}{llrrr}
\toprule
Model & Readout & TP & FN & TPR \\
\midrule
Llama 3.1 8B & Mean diff. & 311 & 9 & 0.972 \\
 & Logistic & 304 & 16 & 0.950 \\
 & RBF-SVM & 309 & 11 & 0.966 \\
 & MLP & 310 & 10 & 0.969 \\
\midrule
Mistral 7B & Mean diff. & 302 & 18 & 0.944 \\
 & Logistic & 296 & 24 & 0.925 \\
 & RBF-SVM & 296 & 24 & 0.925 \\
 & MLP & 302 & 18 & 0.944 \\
\midrule
Qwen 2.5 7B & Mean diff. & 313 & 7 & 0.978 \\
 & Logistic & 295 & 25 & 0.922 \\
 & RBF-SVM & 308 & 12 & 0.963 \\
 & MLP & 303 & 17 & 0.947 \\
\midrule
Mistral Small 24B & Mean diff. & 303 & 17 & 0.947 \\
 & Logistic & 303 & 17 & 0.947 \\
 & RBF-SVM & 309 & 11 & 0.966 \\
 & MLP & 309 & 11 & 0.966 \\
\midrule
Qwen 2.5 32B & Mean diff. & 307 & 13 & 0.959 \\
 & Logistic & 301 & 19 & 0.941 \\
 & RBF-SVM & 310 & 10 & 0.969 \\
 & MLP & 304 & 16 & 0.950 \\
\bottomrule
\end{tabular}
\par\smallskip
\begin{minipage}{\columnwidth}\footnotesize
All decisions use the full stored threshold and the strict rule $s>\tau$. Equality is negative. These summaries were added after the original results were frozen, without refitting or recalibration.
\end{minipage}
\end{table}

\begin{table*}[t]
\centering
\caption{Reference-calibrated operating points. Each readout transfers its unchanged reference threshold to all four Twin cohorts.}
\label{tab:reference-calibration}
\footnotesize
\setlength{\tabcolsep}{3pt}
\begin{tabular*}{\textwidth}{@{\extracolsep{\fill}}llrcccc@{}}
\toprule
\multicolumn{7}{@{}l}{\textit{Reference validation: six fixed calibrations}} \\
Model & Readout & $\tau$ & TP/FN & FP/TN & TPR [95\% CI] & FBR [95\% CI] \\
\midrule
Llama 3.1 8B & Whitened mean difference & 5.400 & 565/30 & 1/478 & 0.950 [0.929, 0.966] & 0.002 [0.000, 0.012] \\
Llama 3.1 8B & Local covariance & 17.609 & 565/30 & 1/478 & 0.950 [0.929, 0.966] & 0.002 [0.000, 0.012] \\
Mistral 7B & Whitened mean difference & 5.626 & 565/30 & 1/478 & 0.950 [0.929, 0.966] & 0.002 [0.000, 0.012] \\
Mistral 7B & Local covariance & 16.564 & 565/30 & 1/478 & 0.950 [0.929, 0.966] & 0.002 [0.000, 0.012] \\
Qwen 2.5 7B & Whitened mean difference & 5.818 & 565/30 & 1/478 & 0.950 [0.929, 0.966] & 0.002 [0.000, 0.012] \\
Qwen 2.5 7B & Local covariance & 15.326 & 565/30 & 1/478 & 0.950 [0.929, 0.966] & 0.002 [0.000, 0.012] \\
\bottomrule
\end{tabular*}
\par\smallskip
\begin{tabular*}{\textwidth}{@{\extracolsep{\fill}}llcccc@{}}
\toprule
Model & Readout & TP/FN & FP/TN & TPR [95\% CI] & FBR [95\% CI] \\
\midrule
\multicolumn{6}{@{}l}{\textit{Twin-70: 70 pairs}} \\
Llama 3.1 8B & Whitened mean difference & 68/2 & 18/52 & 0.971 [0.901, 0.997] & 0.257 [0.160, 0.376] \\
Llama 3.1 8B & Local covariance & 61/9 & 11/59 & 0.871 [0.770, 0.939] & 0.157 [0.081, 0.264] \\
Mistral 7B & Whitened mean difference & 68/2 & 14/56 & 0.971 [0.901, 0.997] & 0.200 [0.114, 0.313] \\
Mistral 7B & Local covariance & 66/4 & 9/61 & 0.943 [0.860, 0.984] & 0.129 [0.061, 0.230] \\
Qwen 2.5 7B & Whitened mean difference & 66/4 & 11/59 & 0.943 [0.860, 0.984] & 0.157 [0.081, 0.264] \\
Qwen 2.5 7B & Local covariance & 64/6 & 11/59 & 0.914 [0.823, 0.968] & 0.157 [0.081, 0.264] \\
\midrule
\multicolumn{6}{@{}l}{\textit{Twin-120: 120 pairs}} \\
Llama 3.1 8B & Whitened mean difference & 104/16 & 16/104 & 0.867 [0.793, 0.922] & 0.133 [0.078, 0.207] \\
Llama 3.1 8B & Local covariance & 86/34 & 6/114 & 0.717 [0.627, 0.795] & 0.050 [0.019, 0.106] \\
Mistral 7B & Whitened mean difference & 106/14 & 14/106 & 0.883 [0.812, 0.935] & 0.117 [0.065, 0.188] \\
Mistral 7B & Local covariance & 99/21 & 8/112 & 0.825 [0.745, 0.888] & 0.067 [0.029, 0.127] \\
Qwen 2.5 7B & Whitened mean difference & 100/20 & 7/113 & 0.833 [0.754, 0.895] & 0.058 [0.024, 0.116] \\
Qwen 2.5 7B & Local covariance & 97/23 & 10/110 & 0.808 [0.726, 0.874] & 0.083 [0.041, 0.148] \\
\midrule
\multicolumn{6}{@{}l}{\textit{Twin-217: 217 pairs}} \\
Llama 3.1 8B & Whitened mean difference & 185/32 & 16/201 & 0.853 [0.798, 0.897] & 0.074 [0.043, 0.117] \\
Llama 3.1 8B & Local covariance & 122/95 & 12/205 & 0.562 [0.493, 0.629] & 0.055 [0.029, 0.095] \\
Mistral 7B & Whitened mean difference & 185/32 & 16/201 & 0.853 [0.798, 0.897] & 0.074 [0.043, 0.117] \\
Mistral 7B & Local covariance & 147/70 & 11/206 & 0.677 [0.611, 0.739] & 0.051 [0.026, 0.089] \\
Qwen 2.5 7B & Whitened mean difference & 175/42 & 11/206 & 0.806 [0.748, 0.857] & 0.051 [0.026, 0.089] \\
Qwen 2.5 7B & Local covariance & 151/66 & 13/204 & 0.696 [0.630, 0.756] & 0.060 [0.032, 0.100] \\
\midrule
\multicolumn{6}{@{}l}{\textit{Twin-43: 43 pairs}} \\
Llama 3.1 8B & Whitened mean difference & 39/4 & 6/37 & 0.907 [0.779, 0.974] & 0.140 [0.053, 0.279] \\
Llama 3.1 8B & Local covariance & 33/10 & 5/38 & 0.767 [0.614, 0.882] & 0.116 [0.039, 0.251] \\
Mistral 7B & Whitened mean difference & 38/5 & 6/37 & 0.884 [0.749, 0.961] & 0.140 [0.053, 0.279] \\
Mistral 7B & Local covariance & 34/9 & 4/39 & 0.791 [0.640, 0.900] & 0.093 [0.026, 0.221] \\
Qwen 2.5 7B & Whitened mean difference & 35/8 & 4/39 & 0.814 [0.666, 0.916] & 0.093 [0.026, 0.221] \\
Qwen 2.5 7B & Local covariance & 33/10 & 4/39 & 0.767 [0.614, 0.882] & 0.093 [0.026, 0.221] \\
\bottomrule
\end{tabular*}
\par\smallskip
\raggedright
Each threshold is the linearly interpolated fifth percentile of the 595 harmful reference-validation scores.
The reference split additionally contains 479 harmless requests. Its rates are in-calibration diagnostics.
Target decisions use $s>\tau$, with equality assigned harmless. Brackets give two-sided 95\% Clopper--Pearson intervals
conditional on the fitted scores and fixed threshold. Displayed thresholds are rounded, while decisions use their full precision.
Llama~3.1 8B and Qwen~2.5 7B use the refusal-modified checkpoints specified in Section~\ref{sec:setup}.
\end{table*}

\subsection{Reference-calibrated operating points}
\label{app:reference-calibration}

This supplementary analysis retains the existing whitened mean-difference
and local-covariance fits on 1,675 harmful and 1,333 harmless reference-fit
requests. It scores the disjoint validation split of 595 harmful and 479
harmless requests. These strong reference-fitted candidates were selected
to challenge the transfer finding, informed by their already observed matched
ranking. This target-informed selection may yield optimistic performance
estimates for new cohorts. The calibration rule and complete six-readout,
four-cohort matrix were fixed before the new operating-point results.

For sorted harmful validation scores, the threshold linearly interpolates
at zero-based position \((595-1)\times0.05\). The decision is strictly
\(s>\tau\), assigning equality to the harmless class. Each threshold is
shared across the four target cohorts. No validation-based sign change,
parameter search, target-label calibration, or row selection is introduced.
The target-score vectors are bit-identical to the corresponding frozen scores.

Table~\ref{tab:reference-calibration} reports all calibrations and target
cells. Intervals are two-sided 95\% Clopper--Pearson intervals, conditional
on the fitted scores and realized thresholds. They do not include uncertainty
from repeating calibration. The reproduction artifact exposes the numerical
inputs and an independent reconstruction of quantiles, counts, and intervals.

For the operating counterconditions in Appendix~\ref{app:wall-counterconditions},
all six Twin-70 FBR upper bounds exceed \(0.20\) (\(0.230\)--\(0.376\)).
On Twin-217, the Llama and Mistral whitened readouts have TPR lower bounds
of \(0.798\), below \(0.80\) despite an FBR upper bound of \(0.117\).

\subsection{Central overlap and secondary operating points}
\label{app:descriptive-completeness}

Table~\ref{tab:central-overlap} gives the interval intersection-over-union
and strict median crossovers defined in Section~\ref{sec:problem}.
Table~\ref{tab:secondary-operating} gives the primary source-threshold
operation on every secondary cohort. All rows retain the complete cohorts
and the original score orientation.

\begin{table*}[t]
\centering
\caption{Central overlap and median crossovers on Twin-70.}
\label{tab:central-overlap}
\footnotesize
\setlength{\tabcolsep}{4pt}
\begin{tabular*}{\textwidth}{@{\extracolsep{\fill}}lccc@{}}
\toprule
Model & Central interval overlap & Harmless above harmful median & Harmful below harmless median \\
\midrule
Llama 3.1 8B & 0.543 & 14/70 (0.200) & 17/70 (0.243) \\
Mistral 7B & 0.481 & 11/70 (0.157) & 12/70 (0.171) \\
Qwen 2.5 7B & 0.326 & 3/70 (0.043) & 11/70 (0.157) \\
Mistral Small 24B & 0.350 & 3/70 (0.043) & 9/70 (0.129) \\
Qwen 2.5 32B & 0.258 & 2/70 (0.029) & 4/70 (0.057) \\
\bottomrule
\end{tabular*}
\par\smallskip
\raggedright
Central overlap is the intersection-over-union length of the harmful and harmless 10th--90th percentile intervals. Crossovers use strict inequalities and linear medians. Equality is excluded from the numerators. All rows contain 70 harmful and 70 harmless requests. These are descriptive quantities, not operating thresholds.
\end{table*}

\begin{table*}[t]
\centering
\caption{Transferred primary operating points on the secondary cohorts.}
\label{tab:secondary-operating}
\footnotesize
\setlength{\tabcolsep}{4pt}
\begin{tabular*}{\textwidth}{@{\extracolsep{\fill}}llrcccc@{}}
\toprule
Cohort & Model & $\tau$ & TP/FN & FP/TN & TPR & FBR \\
\midrule
Twin-120 & Llama 3.1 8B & 8.871 & 120/0 & 119/1 & 1.000 & 0.992 \\
Twin-120 & Mistral 7B & 66.383 & 120/0 & 119/1 & 1.000 & 0.992 \\
Twin-120 & Qwen 2.5 7B & 11.657 & 106/14 & 74/46 & 0.883 & 0.617 \\
Twin-120 & Mistral Small 24B & 40.746 & 120/0 & 115/5 & 1.000 & 0.958 \\
Twin-120 & Qwen 2.5 32B & 10.508 & 107/13 & 61/59 & 0.892 & 0.508 \\
Twin-217 & Llama 3.1 8B & 8.871 & 217/0 & 216/1 & 1.000 & 0.995 \\
Twin-217 & Mistral 7B & 66.383 & 217/0 & 215/2 & 1.000 & 0.991 \\
Twin-217 & Qwen 2.5 7B & 11.657 & 193/24 & 135/82 & 0.889 & 0.622 \\
Twin-217 & Mistral Small 24B & 40.746 & 216/1 & 209/8 & 0.995 & 0.963 \\
Twin-217 & Qwen 2.5 32B & 10.508 & 191/26 & 108/109 & 0.880 & 0.498 \\
Twin-43 & Llama 3.1 8B & 8.871 & 43/0 & 43/0 & 1.000 & 1.000 \\
Twin-43 & Mistral 7B & 66.383 & 43/0 & 43/0 & 1.000 & 1.000 \\
Twin-43 & Qwen 2.5 7B & 11.657 & 40/3 & 33/10 & 0.930 & 0.767 \\
Twin-43 & Mistral Small 24B & 40.746 & 43/0 & 43/0 & 1.000 & 1.000 \\
Twin-43 & Qwen 2.5 32B & 10.508 & 36/7 & 20/23 & 0.837 & 0.465 \\
\bottomrule
\end{tabular*}
\par\smallskip
\raggedright
The source-derived threshold is unchanged across cohorts. A request is classified harmful only when $s>\tau$. Displayed thresholds are rounded. Decisions use full precision. TP/FN and FP/TN are absolute counts, TPR is harmful recall, and FBR is the harmless false-block rate.
\end{table*}

\subsection{Sequence and transition controls}
\label{app:sequence-controls}

Table~\ref{tab:sequence-controls} reports both controls defined in
Appendix~\ref{app:representation-detail}, alongside their respective
baselines. These comparisons concern representations trained with matched
labels. Their CV-BTE also uses thresholds selected on matched training
folds. The artifact provides the individual scores, fold assignments,
thresholds, paired rankings and confusion counts for all 48 score cells.

\begin{table*}[t]
\centering
\caption{Sequence and transition controls across all four Twin cohorts.}
\label{tab:sequence-controls}
\footnotesize
\setlength{\tabcolsep}{4pt}
\begin{tabular*}{\textwidth}{@{\extracolsep{\fill}}llcccccc@{}}
\toprule
Cohort & Model & \multicolumn{2}{c}{AUROC} & \multicolumn{2}{c}{Pair ranking} & \multicolumn{2}{c}{CV-BTE} \\
\midrule
 & & Base & Control & Base & Control & Base & Control \\
\midrule
\multicolumn{8}{@{}l}{\textit{Recurrent sequence: token-order readout and token-shuffle control}} \\
Twin-70 & Llama 3.1 8B & 0.959 & 0.967 & 0.986 & 0.986 & 0.129 & 0.071 \\
Twin-70 & Mistral 7B & 0.969 & 0.948 & 0.986 & 1.000 & 0.093 & 0.100 \\
Twin-70 & Qwen 2.5 7B & 0.954 & 0.930 & 0.971 & 0.971 & 0.107 & 0.200 \\
Twin-120 & Llama 3.1 8B & 0.965 & 0.963 & 1.000 & 0.992 & 0.104 & 0.138 \\
Twin-120 & Mistral 7B & 0.967 & 0.968 & 1.000 & 0.992 & 0.083 & 0.083 \\
Twin-120 & Qwen 2.5 7B & 0.975 & 0.966 & 0.983 & 0.983 & 0.113 & 0.133 \\
Twin-217 & Llama 3.1 8B & 0.969 & 0.961 & 1.000 & 1.000 & 0.094 & 0.090 \\
Twin-217 & Mistral 7B & 0.981 & 0.968 & 1.000 & 0.995 & 0.065 & 0.085 \\
Twin-217 & Qwen 2.5 7B & 0.975 & 0.961 & 0.995 & 0.991 & 0.076 & 0.122 \\
Twin-43 & Llama 3.1 8B & 0.909 & 0.791 & 0.977 & 0.977 & 0.174 & 0.291 \\
Twin-43 & Mistral 7B & 0.944 & 0.880 & 1.000 & 0.977 & 0.140 & 0.163 \\
Twin-43 & Qwen 2.5 7B & 0.915 & 0.916 & 1.000 & 1.000 & 0.105 & 0.186 \\
\midrule
\multicolumn{8}{@{}l}{\textit{Transition asymmetry: directed readout and symmetric control}} \\
Twin-70 & Llama 3.1 8B & 0.829 & 0.831 & 0.829 & 0.857 & 0.243 & 0.307 \\
Twin-70 & Mistral 7B & 0.912 & 0.893 & 0.957 & 0.929 & 0.164 & 0.221 \\
Twin-70 & Qwen 2.5 7B & 0.861 & 0.876 & 0.843 & 0.843 & 0.193 & 0.164 \\
Twin-120 & Llama 3.1 8B & 0.834 & 0.822 & 0.875 & 0.858 & 0.237 & 0.233 \\
Twin-120 & Mistral 7B & 0.833 & 0.852 & 0.850 & 0.850 & 0.254 & 0.233 \\
Twin-120 & Qwen 2.5 7B & 0.885 & 0.873 & 0.908 & 0.908 & 0.179 & 0.200 \\
Twin-217 & Llama 3.1 8B & 0.837 & 0.813 & 0.862 & 0.816 & 0.251 & 0.260 \\
Twin-217 & Mistral 7B & 0.883 & 0.864 & 0.931 & 0.926 & 0.196 & 0.242 \\
Twin-217 & Qwen 2.5 7B & 0.853 & 0.822 & 0.899 & 0.880 & 0.217 & 0.253 \\
Twin-43 & Llama 3.1 8B & 0.796 & 0.835 & 0.767 & 0.884 & 0.279 & 0.279 \\
Twin-43 & Mistral 7B & 0.906 & 0.893 & 0.977 & 0.953 & 0.105 & 0.174 \\
Twin-43 & Qwen 2.5 7B & 0.804 & 0.822 & 0.837 & 0.791 & 0.326 & 0.233 \\
\bottomrule
\end{tabular*}
\par\smallskip
\raggedright
These token-shuffle and symmetric-transition controls differ from the pair-shuffle diagnostic. Pair ranking gives half credit to tied pairs. CV-BTE is the balanced error of the stored pair-grouped held-out decisions, using the existing fold thresholds. The fold thresholds were selected on matched training pairs. All values are point estimates. Scores, correct/wrong/tied pair counts and CV confusion counts accompany the reproduction code.
\end{table*}

\section{Surface-Form and Sensor Analyses}

\subsection{Generation-time sensor}
\label{app:sensor}

The separate generation-time sensor uses the fixed-reference score defined as
the generation-window method and a fixed zero threshold. It evaluates
activations from the generated continuation rather than the prompt-only
source readout. Its references are a separately frozen set of five medoids
for each of 43 categories, giving 215 reference vectors per checkpoint.
Manual-review coverage of this separate medoid set is not established.

Table~\ref{tab:sensor} reports its Twin-70 and XSTest operating
points. These results illustrate another scalar sensor boundary but do not
enter the activation-transfer or Wall assessment.

\begin{table}[!htbp]
\centering
\caption{Operating points of the separate generation-time sensor.}
\label{tab:sensor}
\footnotesize
\setlength{\tabcolsep}{3pt}
\begin{tabular*}{\columnwidth}{@{\extracolsep{\fill}}>{\raggedright\arraybackslash}p{0.30\linewidth}ccc@{}}
\toprule
Metric & \begin{tabular}[c]{@{}c@{}}{Llama 3.1}\\ {8B}\end{tabular} & \begin{tabular}[c]{@{}c@{}}{Mistral}\\ {7B}\end{tabular} & \begin{tabular}[c]{@{}c@{}}{Qwen 2.5}\\ {7B}\end{tabular} \\
\addlinespace[2pt]
\midrule
Twin-70 TPR & 0.900 & 0.986 & 0.971 \\
Twin-70 TP/FN & 63/7 & 69/1 & 68/2 \\
Twin-70 FBR & 0.457 & 0.486 & 0.743 \\
Twin-70 FP/TN & 32/38 & 34/36 & 52/18 \\
XSTest FP/TN & 158/92 & 159/91 & 175/75 \\
XSTest FBR & 0.632 & 0.636 & 0.700 \\
\bottomrule
\end{tabular*}
\par\smallskip
\raggedright All results use the sensor's fixed zero threshold with $s>0$. Equality is negative.
\end{table}

\subsection{Surface-form controls}
\label{app:surface}

\paragraph{Protocol and decision rules}
This case study asks whether a score or threshold decision changes while
the model still correctly identifies the more harmful arm of a pair.

The primary surface-form readout uses Qwen~2.5 32B in bfloat16, zero-indexed
layer 47, and componentwise maximum pooling. Its direction is fitted on the
same broad source fit split as the primary transfer readout. A diagnostic
threshold is then set to the fifth percentile of harmful scores in the
unmodified Twin-151 condition. Its use of target labels makes it a
target-exposed diagnostic rather than a deployable transferred operating
point.

We apply three deterministic transformations:

\begin{itemize}
\item \(T_{\mathrm{space}}\) inserts a space between adjacent non-space
characters.
\item \(T_{\mathrm{underscore}}\) appends an underscore to each alphabetic
Unicode character.
\item \(T_{\mathrm{control}}\) spaces only a fixed conservative set of
function words to control partly for length and tokenization changes.
\end{itemize}

The activation readout retains the first 64 tokens of each transformed
request. Both systematic transformations exceed this window for every arm,
so they change retained token coverage as well as surface form. The
behavioral check instead presents each complete pair in native chat format.
The transformations are evaluated separately. Each of the 151 pairs is
presented once in each A/B order, asking which arm would be more harmful if
fulfilled. Let \(c_{i,AB}\) and \(c_{i,BA}\) indicate correct choices after
accounting for the arm swap. Presentation accuracy is
\(\sum_i(c_{i,AB}+c_{i,BA})/(2N)\). Invalid choices count as incorrect, and
the bootstrap retains both orders of each pair. We additionally report
both-orders pair consistency, \(\sum_i c_{i,AB}c_{i,BA}/N\), as a
descriptive check. It is distinct from score-based pair ranking.

The surface threshold evasion rate (STER) requires all four choices to be
correct: both orders in the unmodified condition and both in the transformed
condition, together with an
unmodified harmful score strictly above the diagnostic threshold. Among these
eligible pairs, STER is the fraction whose transformed harmful score is at or
below the threshold. An empty eligible set has an undefined rate and interval.
STER is a conditional readout statistic, not an attack success rate.

A transformation is classified as directional degradation only when the
unmodified score-based pair-ranking lower bound exceeds \(0.50\), the
transformed presentation-accuracy lower bound is at least \(0.80\), the lower bound on its
change from the unmodified condition is at least \(-0.10\), and the upper
bounds on the AUROC and pair-ranking changes are below zero. It is classified
as material degradation only when the upper bound on the pair-ranking change
is below \(-0.10\).

A separately fitted four-bit (Q4) implementation provides a
numerical-precision comparison. Its scores and target-derived threshold are
not numerically comparable with the bfloat16 readout. The available model
records leave their weight-snapshot relationship unresolved. This comparison
therefore represents two implementation regimes rather than an estimated
effect of reducing precision.

\paragraph{Complete results}

Tables~\ref{tab:surface_behavior} and~\ref{tab:surface_appendix} report
behavioral accuracy and the complete componentwise-maximum surface-form
results, respectively. Each numerical regime uses its own source-trained readout
and its own diagnostic threshold derived from the unmodified Twin-151
harmful scores. The Q4 rows provide an implementation and numerical-precision
comparison under their own score scale.
The presentation-accuracy column uses all 302 choices per condition, including
invalid answers as errors. The adjacent both-orders count instead requires two
correct choices per pair. There are no repeated trials or excluded pairs.
One invalid choice occurs in BF16 unmodified and one in BF16 underscore.
Behavior responses are shared across pooling rules, whereas STER uses each
pooling rule's own scores, threshold, and eligible set. Complete 18-cell STER
counts and intervals are supplied in the text-free artifact. Its standalone
STER intervals use Wilson's method, with no bootstrap draws or bootstrap
denominators.

\begin{table}[!htbp]
\centering
\caption{Qwen~2.5 32B decisions under surface transformations.}
\label{tab:surface_behavior}
\footnotesize
\setlength{\tabcolsep}{3pt}
\begin{tabular*}{\columnwidth}{@{\extracolsep{\fill}}>{\raggedright\arraybackslash}p{0.24\linewidth}ccc@{}}
\toprule
Condition & \begin{tabular}[c]{@{}c@{}}{Presentation}\\ {accuracy}\end{tabular} & \begin{tabular}[c]{@{}c@{}}{Both-orders}\\ {pair accuracy}\end{tabular} & Invalid \\
\midrule
\multicolumn{4}{@{}l}{\textit{BF16}} \\
Unmodified & 295/302 (0.977) & 144/151 (0.954) & 1 \\
$T_{\mathrm{space}}$ & 297/302 (0.983) & 146/151 (0.967) & 0 \\
$T_{\mathrm{underscore}}$ & 295/302 (0.977) & 144/151 (0.954) & 1 \\
$T_{\mathrm{control}}$ & 301/302 (0.997) & 150/151 (0.993) & 0 \\
\midrule
\multicolumn{4}{@{}l}{\textit{Q4}} \\
Unmodified & 297/302 (0.983) & 146/151 (0.967) & 0 \\
$T_{\mathrm{space}}$ & 299/302 (0.990) & 148/151 (0.980) & 0 \\
$T_{\mathrm{underscore}}$ & 297/302 (0.983) & 146/151 (0.967) & 0 \\
$T_{\mathrm{control}}$ & 300/302 (0.993) & 149/151 (0.987) & 0 \\
\bottomrule
\end{tabular*}
\par\smallskip
\raggedright Cells give numerator/denominator (accuracy). Presentation accuracy counts individual A/B presentations. Both-orders pair accuracy requires both presentations to be correct. Invalid presentations count as incorrect. The behavior test is shared by all pooling rules.
\end{table}

\begin{table}[!htbp]
\centering
\caption{Qwen~2.5 32B readout under surface transformations.}
\label{tab:surface_appendix}
\footnotesize
\setlength{\tabcolsep}{3pt}
\begin{tabular*}{\columnwidth}{@{\extracolsep{\fill}}>{\raggedright\arraybackslash}p{0.21\linewidth}cccc@{}}
\toprule
Metric & Unmodified & $T_{\mathrm{space}}$ & $T_{\mathrm{underscore}}$ & $T_{\mathrm{control}}$ \\
\midrule
\multicolumn{5}{@{}l}{\textit{BF16}} \\
AUROC & 0.824 & 0.732 & 0.721 & 0.821 \\
Pair rank & 0.874 & 0.795 & 0.728 & 0.887 \\
Score C/W/T & 132/19/0 & 114/25/12 & 104/35/12 & 134/17/0 \\
TPR & 0.947 & 0.583 & 0.616 & 0.967 \\
STER & \begin{tabular}[c]{@{}c@{}}{not}\\ {applicable}\end{tabular} & \begin{tabular}[c]{@{}c@{}}{50/135}\\ {(0.370)}\end{tabular} & \begin{tabular}[c]{@{}c@{}}{45/135}\\ {(0.333)}\end{tabular} & \begin{tabular}[c]{@{}c@{}}{1/138}\\ {(0.007)}\end{tabular} \\
\addlinespace[2pt]
Assessment & baseline & directional & directional & control \\
\midrule
\multicolumn{5}{@{}l}{\textit{Q4}} \\
AUROC & 0.828 & 0.751 & 0.715 & 0.825 \\
Pair rank & 0.901 & 0.801 & 0.762 & 0.894 \\
Score C/W/T & 136/15/0 & 115/24/12 & 109/30/12 & 135/16/0 \\
TPR & 0.947 & 0.603 & 0.675 & 0.974 \\
STER & \begin{tabular}[c]{@{}c@{}}{not}\\ {applicable}\end{tabular} & \begin{tabular}[c]{@{}c@{}}{51/138}\\ {(0.370)}\end{tabular} & \begin{tabular}[c]{@{}c@{}}{41/138}\\ {(0.297)}\end{tabular} & \begin{tabular}[c]{@{}c@{}}{1/140}\\ {(0.007)}\end{tabular} \\
\addlinespace[2pt]
Assessment & baseline & directional & directional & control \\
\bottomrule
\end{tabular*}
\par\smallskip
\raggedright Componentwise-maximum pooling with a separately fitted diagnostic threshold in each regime. Score C/W/T gives correct, wrong, and tied score rankings. STER gives crossings over eligible original-positive harmful arms (rate), requiring both orders correct in both conditions. It does not measure harmful compliance. Directional denotes directional, not material, degradation.
\end{table}

Table~\ref{tab:surface_intervals} gives the frozen intervals that determine
the primary BF16 surface-form classifications.
The surface criteria remain those defined above. In
particular, a negative pair-ranking change can establish directional
degradation without meeting the stricter material-degradation margin. The
unmodified BF16 pair-ranking lower bound is
\SurfaceBFOriginalPairRankingLowerBound{}, above the required \(0.50\).

\begin{table}[!htbp]
\centering
\caption{Uncertainty for Qwen~2.5 32B BF16 surface-form diagnostics.}
\label{tab:surface_intervals}
\footnotesize
\setlength{\tabcolsep}{3pt}
\begin{tabular*}{\columnwidth}{@{\extracolsep{\fill}}>{\raggedright\arraybackslash}p{0.30\linewidth}cc@{}}
\toprule
Metric & $T_{\mathrm{space}}$ & $T_{\mathrm{underscore}}$ \\
\midrule
Presentation acc. & \begin{tabular}[c]{@{}c@{}}{0.983}\\ {[0.967, 0.997]}\end{tabular} & \begin{tabular}[c]{@{}c@{}}{0.977}\\ {[0.960, 0.990]}\end{tabular} \\
\addlinespace[2pt]
$\Delta$acc. & \begin{tabular}[c]{@{}c@{}}{0.007}\\ {[$-0.010$, 0.026]}\end{tabular} & \begin{tabular}[c]{@{}c@{}}{0.000}\\ {[$-0.017$, 0.017]}\end{tabular} \\
\addlinespace[2pt]
$\Delta$AUROC & \begin{tabular}[c]{@{}c@{}}{$-0.093$}\\ {[$-0.138$, $-0.046$]}\end{tabular} & \begin{tabular}[c]{@{}c@{}}{$-0.103$}\\ {[$-0.149$, $-0.059$]}\end{tabular} \\
\addlinespace[2pt]
$\Delta$pair rank & \begin{tabular}[c]{@{}c@{}}{$-0.079$}\\ {[$-0.149$, $-0.007$]}\end{tabular} & \begin{tabular}[c]{@{}c@{}}{$-0.146$}\\ {[$-0.219$, $-0.073$]}\end{tabular} \\
\addlinespace[2pt]
STER & \begin{tabular}[c]{@{}c@{}}{0.370}\\ {[0.294, 0.454]}\end{tabular} & \begin{tabular}[c]{@{}c@{}}{0.333}\\ {[0.259, 0.416]}\end{tabular} \\
\addlinespace[2pt]
\bottomrule
\end{tabular*}
\par\smallskip
\raggedright Brackets give 95\% intervals. The readout uses componentwise-maximum pooling. Presentation accuracy averages both A/B orders. Its intervals and paired changes keep orders within pairs. STER uses Wilson intervals. Changes are transformed minus unmodified. Both transformations satisfy directional, not material, degradation.
\end{table}

\section{Blinded Second-Judge Diagnostics}
\label{app:second-judge}

Tables~\ref{tab:second_judge} and~\ref{tab:second_judge_sensitivity} report
the automated judge audit and label-sensitivity analysis. The
consensus subset treats an uncertain isolated-arm label as an
abstention and removes only direct binary contradictions. The strict subset
also removes pairs with an uncertain, blocked, incomplete, or invalid arm
response. All retained rows preserve our frozen manual labels. The maximum
absolute changes are taken across \SecondJudgeReadoutCount{} readouts and
four cohorts separately for each subset. Swapped-order pair fields are descriptive
diagnostics and never determine membership.

\begingroup
\makeatletter
\renewenvironment{table}[1][]{%
  \par\addvspace{\intextsep}\noindent
  \begin{minipage}{\columnwidth}\def\@captype{table}%
}{\end{minipage}\par\addvspace{\intextsep}}
\makeatother
\begin{table}[!htbp]
\centering
\caption{Blinded model-based second-judge agreement.}
\label{tab:second_judge}
\footnotesize
\setlength{\tabcolsep}{3pt}
\begin{tabular*}{\columnwidth}{@{\extracolsep{\fill}}>{\raggedright\arraybackslash}p{0.42\linewidth}cc@{}}
\toprule
Measure & \begin{tabular}[c]{@{}c@{}}{Estimate}\\ {[95\% CI]}\end{tabular} & Denominator \\
\midrule
\multicolumn{3}{@{}l}{\textit{Isolated-arm label audit}} \\
Raw agreement with manual labels & \begin{tabular}[c]{@{}c@{}}{\SecondJudgeRawAgreement{}}\\ {[\SecondJudgeRawAgreementLower{}, \SecondJudgeRawAgreementUpper{}]}\end{tabular} & \begin{tabular}[c]{@{}c@{}}{\SecondJudgeArmCount{}}\\ {arms}\end{tabular} \\
Binary Cohen's $\kappa$ & \begin{tabular}[c]{@{}c@{}}{\SecondJudgeBinaryKappa{}}\\ {[\SecondJudgeBinaryKappaLower{}, \SecondJudgeBinaryKappaUpper{}]}\end{tabular} & \begin{tabular}[c]{@{}c@{}}{866}\\ {binary}\\ {responses}\end{tabular} \\
Three-class Cohen's $\kappa$ & \begin{tabular}[c]{@{}c@{}}{\SecondJudgeThreeClassKappa{}}\\ {[\SecondJudgeThreeClassKappaLower{}, \SecondJudgeThreeClassKappaUpper{}]}\end{tabular} & \begin{tabular}[c]{@{}c@{}}{\SecondJudgeArmCount{}}\\ {arms}\end{tabular} \\
Harmful / harmless agreement & \SecondJudgeHarmfulAgreement{} / \SecondJudgeHarmlessAgreement{} & \begin{tabular}[c]{@{}c@{}}{450 / 450}\\ {arms}\end{tabular} \\
Nonbinary / blocked rate & \SecondJudgeUncertainRate{} / \SecondJudgeBlockedRate{} & \begin{tabular}[c]{@{}c@{}}{\SecondJudgeArmCount{}}\\ {arms}\end{tabular} \\
\midrule
\multicolumn{3}{@{}l}{\textit{Swapped-order pair assessment}} \\
Pair validity / harmful orientation & \SecondJudgePairValidityAgreement{} / \SecondJudgeOrientationAgreement{} & \begin{tabular}[c]{@{}c@{}}{\SecondJudgePresentationCount{}}\\ {presentations}\end{tabular} \\
Exactly one harmful / same topic & \SecondJudgeExactlyOneAgreement{} / \SecondJudgeSameTopicAgreement{} & \begin{tabular}[c]{@{}c@{}}{\SecondJudgePresentationCount{}}\\ {presentations}\end{tabular} \\
Same request frame & \SecondJudgeSameFrameAgreement{} & \begin{tabular}[c]{@{}c@{}}{\SecondJudgePresentationCount{}}\\ {presentations}\end{tabular} \\
All-field / orientation swap consistency & \SecondJudgeAllFieldSwapConsistency{} / \SecondJudgeOrientationSwapConsistency{} & \begin{tabular}[c]{@{}c@{}}{\SecondJudgePairCount{}}\\ {pairs}\end{tabular} \\
\bottomrule
\end{tabular*}
\par\smallskip
\raggedright The first three measures report pair-clustered 95\% bootstrap intervals. Other intervals are not reported. The 34 nonbinary records comprise 29 explicit uncertain responses and five blocked calls recorded as uncertain. Blocked calls are included in both rates. The swapped-order pair assessment is diagnostic and does not select either sensitivity subset.
\end{table}

\endgroup
\begin{table}[!htbp]
\centering
\caption{Prespecified isolated-arm label-sensitivity analysis.}
\label{tab:second_judge_sensitivity}
\footnotesize
\setlength{\tabcolsep}{3pt}
\begin{tabular*}{\columnwidth}{@{\extracolsep{\fill}}>{\raggedright\arraybackslash}p{0.39\linewidth}ccc@{}}
\toprule
Measure & Full & Consensus & Strict \\
\midrule
Retained pairs & 450 & \SecondJudgeConsensusPairCount{} & \SecondJudgeStrictPairCount{} \\
Maximum $|\Delta|$, Twin AUROC & -- & \SecondJudgeConsensusMaxTwinShift{} & \SecondJudgeStrictMaxTwinShift{} \\
Maximum $|\Delta|$, pair ranking & -- & \SecondJudgeConsensusMaxPairShift{} & \SecondJudgeStrictMaxPairShift{} \\
Maximum $|\Delta|$, CV-BTE & -- & \SecondJudgeConsensusMaxCVBTEShift{} & \SecondJudgeStrictMaxCVBTEShift{} \\
Maximum $|\Delta|$, fixed TPR & -- & \SecondJudgeConsensusMaxTPRShift{} & \SecondJudgeStrictMaxTPRShift{} \\
Maximum $|\Delta|$, fixed FBR & -- & \SecondJudgeConsensusMaxFBRShift{} & \SecondJudgeStrictMaxFBRShift{} \\
\bottomrule
\end{tabular*}
\par\smallskip
\raggedright Retained pairs span four cohorts. Maximum absolute changes compare each subset with the full cohort over 28 cells per subset. Full-cohort differences are zero by definition and shown as --. Confidence intervals are not reported for these maxima.
\end{table}

\section{Matched-Cohort Construction Detail}
\label{app:cohort-construction}

\subsection{Operational label and matching rubric}

During manual review, we labeled the expressed request rather than a
hypothesized latent user intent. A harmful arm requests an action whose
fulfillment would materially enable, facilitate, or carry out harm. Dangerous
vocabulary or harmful
material that is being analyzed, labeled, translated, moderated, or inspected
defensively is not sufficient by itself. A harmless arm instead requests
legitimate analytical, moderation, research, administrative, preventive, or
authorized defensive work. Authorization is considered only when it is
expressed in the request. If the requested action, target, authorization, or
likely effect cannot be determined, the arm is uncertain rather than forced
into a binary class. Dual-use subject matter therefore does not determine the
label on its own.

A candidate pair also had to preserve the same concrete topic and a comparable
request frame while making the requested harmful versus legitimate action
clear. Review checked directness, structural match, surface markers, missing
context, duplicates, pair validity, and arm orientation. The available actions
were Accept, Revise, Exclude, and Uncertain. Nonaccept decisions required a
reason and could trigger a recorded replacement. Activation, probe, guard,
sensor, automated-judge, and manuscript results were not used to select active
cohort membership.

\subsection{Construction and review records}

We manually reviewed and labeled all candidate pairs under the rubric above.
The four active cohorts contain 450 accepted pairs. The final release check
confirmed their labels and one arm-orientation correction. The reproduction
artifact reports aggregate review and construction counts, selection rules,
and final cohort sizes.

For the primary Twin-70 cohort, blinded review covered two ranked candidates
for each of 70 positions. Of 140 candidates, 132 were marked Accept, seven
Revise, and one Exclude. The fixed rank rule selected 66 rank-1 and four
rank-2 candidates. The original review sheet omitted separate fields for
request-form match, missing context or ambiguity, and overall pair validity
specified in the frozen protocol. All 140 candidates received completed
intent-clarity, directness, structure, duplication, and final-decision entries.

\newpage
\paragraph{Twin-70 generator settings}
The locally hosted Qwen~2.5 32B Instruct generator in \texttt{Q4\_K\_M}
quantization ran serially
through \texttt{llama-server}, with temperature 0.62, top-\(p\) 0.90, at
most 420 generated tokens, seed base 72,719,000, and at most 24 base attempts
per candidate rank. Each draft allowed up to three length corrections at
temperature 0.20, retaining 20--38 words per arm. The construction manifest
records the prompt template, model-file hashes, software version, seed
formula, attempts, rejection reasons, and output hashes. The other cohorts'
preserved records do not provide equally complete generator identities and
decoding settings.

\section{Reproducibility and Data Access}
\label{app:reproducibility}

The Open Science statement specifies the released materials, supported
offline reproduction, and withheld data. The accompanying configurations
record tokenizer versions, data splits, preprocessing, activation locations,
pooling rules, fitted directions, thresholds, fold assignments, and
uncertainty procedures. The checkpoint identities below identify the model
inputs needed for extraction, which the text-free package cannot repeat.

\subsection{Checkpoint identities}

The BF16 checkpoints and MiniLM control use the following Hugging Face
repositories. Revision prefixes are shown here, and the reproduction materials provide the complete
immutable hashes.

{\footnotesize
\begin{itemize}
\item Llama~3.1 8B:\\
\path{mlabonne/Meta-Llama-3.1-8B-Instruct-abliterated}\\
revision \texttt{368c8ed94ce4}.
\item Mistral~7B:\\
\path{mistralai/Mistral-7B-Instruct-v0.3}\\
revision \texttt{c170c708c41d}.
\item Qwen~2.5 7B:\\
\path{huihui-ai/Qwen2.5-7B-Instruct-abliterated}\\
revision \texttt{d5e2ee7eacf6}.
\item Mistral Small 24B:\\
\path{mistralai/Mistral-Small-24B-Instruct-2501}\\
revision \texttt{9527884be6e5}.
\item Qwen~2.5 32B:\\
\path{Qwen/Qwen2.5-32B-Instruct}\\
revision \texttt{5ede1c97bbab}.
\item MiniLM text-embedding control:\\
\path{sentence-transformers/all-MiniLM-L6-v2}\\
revision \texttt{1110a243fdf4}.
\end{itemize}
}

The Q4 file hashes identify the four-bit model used in the surface-form
comparison. We could not verify that this conversion originated from the
same weight revision as the BF16 checkpoint. Differences between them
therefore cannot be attributed to quantization alone.

\clearpage
\section{Direct Matched-Label Results}
\label{app:direct-readouts}

Table~\ref{tab:direct-readouts} separates held-out Twin-70 performance from
transfer of the Twin-70 ensembles to the two secondary cohorts. The fitting
procedure is specified in Appendix~\ref{app:representation-detail}.

\begin{table}[!htbp]
\centering
\caption{Direct matched-label classifiers on Twin-70 and cross-cohort transfer.}
\label{tab:direct-readouts}
\footnotesize
\setlength{\tabcolsep}{3pt}
\begin{tabular*}{\columnwidth}{@{\extracolsep{\fill}}lrrrr@{}}
\toprule
Readout & \multicolumn{2}{c}{Twin-70 out of fold} & \multicolumn{2}{c}{Transfer AUROC} \\
 & AUROC & CV-BTE & Twin-120 & Twin-217 \\
\midrule
\multicolumn{5}{@{}l}{\textit{Llama 3.1 8B}} \\
LR & 0.994 & 0.050 & 0.982 & 0.939 \\
RBF-SVM & 0.991 & 0.057 & 0.979 & 0.930 \\
MLP & 0.994 & 0.064 & 0.980 & 0.933 \\
\midrule
\multicolumn{5}{@{}l}{\textit{Mistral 7B}} \\
LR & 0.995 & 0.036 & 0.978 & 0.973 \\
RBF-SVM & 0.990 & 0.064 & 0.981 & 0.969 \\
MLP & 0.994 & 0.057 & 0.977 & 0.968 \\
\midrule
\multicolumn{5}{@{}l}{\textit{Qwen 2.5 7B}} \\
LR & 0.986 & 0.086 & 0.972 & 0.949 \\
RBF-SVM & 0.980 & 0.114 & 0.977 & 0.944 \\
MLP & 0.982 & 0.071 & 0.975 & 0.949 \\
\bottomrule
\end{tabular*}
\par\smallskip
\raggedright
LR denotes logistic regression. Twin-70 scores use pair-grouped nested cross-validation.
CV-BTE selects thresholds on the corresponding training folds and is diagnostic.
Transfer scores average the five Twin-70 outer-fold models without target-label fitting, selection, scaling, or calibration.
The artifact's \texttt{evidence/direct\_readouts.csv} records all values, 95\% intervals, and analysis identifiers from the frozen aggregate index.
Llama~3.1 8B and Qwen~2.5 7B use the refusal-modified checkpoints specified in Section~\ref{sec:setup}.
\end{table}

\end{document}